\documentclass[floatfix,amsmath,amssymb,aps,twocolumn,superscriptaddress,prb]{revtex4-1}

\usepackage{H1}

\def\KeyWord#1{$\backslash$\IfColor{$\!\!$\textRed{#1}\textBlack}{#1}$\!\!$}

\newcommand{\oS}{\hat{ \bf {S} } }
\newcommand{\oW}{\hat{ \bf {W} } }
\newcommand{\os}{\hat{ \bf {s} } }
\newcommand{\oh}{\hat{\bf{h}}}
\newcommand{\oQ}{\hat{ \mac{Q} }}
\newcommand{\gz}{$\mathbb{Z}_2$ }
\newcommand{\zp}{$\mathbb{Z}_p$ }
\newcommand{\wM}{\widetilde{M} }
\newcommand{\wa}{\widetilde{a} }

\def\AWW{CAWW }

\newcommand{\suw}{SU(2)$_k$ Walker-Wang }

\graphicspath{{Figures/}}

\begin{document}

\title{Phase transitions in three-dimensional topological lattice models with surface anyons}

\author{F.J.~Burnell}
\affiliation{All Souls College, Oxford, United Kingdom}
\affiliation{Rudolf Peierls Centre for Theoretical Physics, 1 Keble Road, Oxford,
OX1 3NP, United Kingdom}
\author{C.W.~von Keyserlingk}
\author{S.H.~Simon}
\affiliation{Rudolf Peierls Centre for Theoretical Physics, 1 Keble Road, Oxford,
OX1 3NP, United Kingdom}
\date{\today}

\begin{abstract}
We study the phase diagrams of a family of 3D ``Walker-Wang" type lattice models, which are not topologically ordered but have deconfined anyonic excitations confined to their surfaces.   We add a perturbation (analogous  to that which drives the confining transition in \zp lattice gauge theories) to the Walker-Wang Hamiltonians, driving a transition in which all or some of the variables associated with the loop gas or string-net ground states of these models become confined.  We show that in many cases the location and nature of the phase transitions involved is exactly that of a generalized \zp lattice gauge theory, and use this to deduce the basic structure of the phase diagram.  We further show that the relationship between the phases on opposite sides of the transition is fundamentally different than in conventional gauge theories: in the Walker-Wang case, the number of species of excitations that are deconfined in the bulk can {\it increase} across a transition that confines only some of the species of loops or string-nets.  The analogue of the confining transition in the Walker-Wang models can therefore lead to bulk deconfinement and topological order.
\end{abstract}

%\pacs{03.75.Ss, 71.10.Ca, 67.85.-d}

\maketitle

\include{epsfx}

\section{Introduction}
Critical phenomena beyond the standard Landau-Ginzburg paradigm have been an active area of interest in condensed matter physics over the past decade.  One area that has drawn significant interest of late is the subject of transitions between phases of different topological order.  \cite{VidalToric,Gilsetal,VidalFibb,TSBLong,WenBarkeshli,WenBarkeshliLong,WB2,TSBShort,TSBLong}

By definition, topological order cannot be diagnosed by any local order parameter\cite{Wen}. Typical hallmarks of this type of order include a finite ground-state degeneracy that changes with the topology of the manifold, topological entanglement entropy\cite{LWTO}, and excitations with anyonic statistics.  For phase transitions that alter the topological order, there is no known analogue of the Landau-Ginzburg approach that correctly describes the critical properties.\footnote{ Kitaev\cite{KitaevUnpublished} has developed a framework similar to the Landau-Ginzburg approach for these systems; however in the form known to the authors this framework necessarily yields mean-field critical exponents.}  However, in 2D there are a number of examples where these transitions are well understood\cite{SlingerlandBais,SBPRL,SBPRL2,VidalToric,Gilsetal,TSBLong,WenBarkeshli,WenBarkeshliLong,WB2}.  Here we will extend this body of work to investigate the phase diagram of a family of 3D models\cite{WW} with interesting topological properties.

In discussing the phase diagram of a topologically ordered system, there are two elements of interest: the nature of the phase transitions, and the topological orders of the phases themselves.  Early progress on the former front was made in Refs. \onlinecite{WegnerJMP12,FradkinShenker} where, amongst other things, it was shown that the confining transition of $\mathbb{Z}_2$ gauge theory is in the Ising universality class.  (A body of later numerical work has followed up on this result\cite{CastelnovoChamon,TrebstTC,VidalToric}, confirming their analytical description of the phase diagram). Much later, Refs. \onlinecite{TSBShort,TSBLong} used similar methods to show that a large family of transitions in 2D Levin-Wen models\cite{LW} fall into the Ising or Potts universality class\footnote{Not all transitions in 2D topologically ordered systems admit such a mapping, however, as the work of Ref's \onlinecite{Gilsetal,VidalFibb} highlights.}.

On the second front,  Slingerland and Bais\cite{SlingerlandBais,SBPRL,SBPRL2} provided a comprehensive criterion -- which we use throughout this work -- for which topological orders can be  connected by a (potentially second-order) phase transition: two phases can be ``naturally" connected by a direct phase transition if one of the phases can arise as the result of forming a condensate in the other.  Two familiar examples are the Higgs transition (Bose condensation of charges), and the confining transitions (Bose condensation of vortices or vortex loops) familiar in gauge theories.  Beyond this, however, their method gives a useful paradigm in 2D for studying condensation transitions in any anyon model. 
 It is worth emphasizing that the problem is not trivial: the low-energy excitations in the condensed phase need not be simply a subset of those of the uncondensed phase.  

In the present work, we will address both the question of the nature of the phase transition, and the relationship between the topological orders before and after condensation, in a family of 3D Walker-Wang\cite{WW}-type models.  
These models 
 are not topologically ordered in the conventional sense\cite{WWUs}: on a system with no boundaries, they always have a unique ground state, and admit no deconfined point-like excitations.  However, on a 3D system with a 2D boundary, they exhibit surface states with chiral topological order (analogous to that of a fractional quantum Hall system) and deconfined anyonic excitations.  Commensurate with these topological characteristics, their ground state wave-functions are 3D cousins of the loop gas or string-net ground states of Levin-Wen models.  We refer to this set of characteristics as {\it surface topological order}, to contrast it with {\it bulk topological order}, in which there is a topological ground-state degeneracy on a system with periodic boundary conditions (i.e. in the absence of surfaces).  This distinction is important: unlike bulk topological order, a system's surface topological order can change without a bulk phase transition.

Though the methodology we use here is similar to that employed in Ref. \onlinecite{TSBLong} to study condensation transitions in Levin-Wen models (the 2D cousins of our 3D models), there are some striking differences between the 2D and 3D systems.  
Notably, in many cases these models admit condensation transitions from phases with surface topological order to phases with 3D bulk topological order. This is in stark contrast to the situation in 3D gauge theories: the transitions we describe are analogous to confining transitions, involving the proliferation of vortex loops; in a gauge theory these can only decrease the topological ground-state degeneracy, and can never produce a phase with bulk topological order from one with none.  (This is because the Higgs transition necessarily reduces the size of the gauge group; for a gapped system the topological ground-state degeneracy vanishes only if the gauge group is trivial).

Our key results are as follows.  We identify a family of condensation transitions which have the same low energy theory as confining transitions in a Potts gauge theory.  These transitions have been studied both numerically\cite{PottsGaugeNumerics,PottsGaugeNumerics,BaigPhysLettB207} and theoretically\cite{PottsGauge1,PottsGauge2}, and are generically first-order.  Our main interest is in exposing the relationship between the pairs of 3D phases related by condensation. We give examples of condensation transitions between Walker-Wang models of trivial bulk topological order and 1) the trivial phase, which has neither bulk nor surface topological order;  2) another Walker-Wang model with only surface topological order; 3) a discrete gauge theory, with only bulk topological order; and 4) a phase with both bulk and surface topological order.  We will show that whenever bulk topological order emerges, the phase also has deconfined point-like excitations, which can be either bosons or fermions, depending on the nature of the condensed phase.  In contrast, in scenarios 2) and 4), we find that if the lattice has boundaries, the models have chiral anyonic excitations confined to these surfaces.  Scenario 4) thus describes a phase with bosonic or fermionic point particles deconfined in the bulk, as well as anyons confined to the surface.

We begin in Sect. \ref{PottsSec} by introducing the Potts gauge theory, which is a useful stepping stone for understanding the Walker-Wang models which we introduce in Sect. \ref{AWWSec}.  We first discuss condensation transitions in the Abelian Walker-Wang models, whose Hamiltonians differ from those of the Potts gauge models only by additional phase factors. These models admit transitions in all of the cases (1-4) above with a minimum of complexity.  

Having understood the phase diagram of these Abelian models, we turn our attention to the non-Abelian case. We begin in Sect. \ref{s:SU(2)} by discussing a family of models with the surface topological order of SU(2)$_k$ Chern-Simons theories.  We show that these exhibit transitions analogous to those of an Ising gauge theory, as well as other types of transitions which do not have such simple analogues.  In Sect. \ref{s:gencat} we describe in more technical language the status of transitions in general Walker-Wang models, giving a general prescription for determining the topological order of the condensed phase.  
\section{Review: condensation in Abelian lattice gauge theories} \label{PottsSec}

We begin by examining condensation transitions in `Potts gauge theories'\cite{PottsGauge1} ( $\integ_{p}^\text{Potts}$ models, for short), which are variations of the well-known discrete $\mathbb{Z}_p$ gauge theories, and which give a natural point of departure from which to discuss transitions in the related Walker-Wang models.\footnote{ These differ from the conventional \zp gauge theories by virtue of the fact that the energy penalty for any magnetic flux $\phi$ is the same, rather than proportional to $\cos \phi$ -- and similarly for electric fluxes.}  We first review the models and their essential features, and then discuss a simple condensation transition which will be our point of reference for understanding all condensation transitions within these models.   We then examine how the topological order changes after the condensation. Expert readers may wish to skim this section for our notation;  we will rely heavily on the framework outlined in this section both to understand the Walker-Wang Hamiltonians in later sections, and to describe the transitions in the resulting phase diagrams.  

\subsection{The \zp `Potts gauge theory' Hamiltonians}

\begin{figure}
\begin{center}
\includegraphics[width=.9\linewidth]{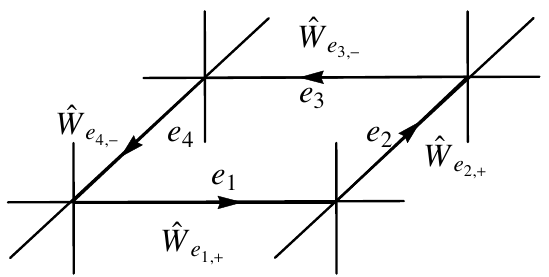}\\
 \caption{  Action of the plaquette operator on a plaquette in the $xy$-plane.  On edges $1$ and $2$, where the arrow points along the $\hat{x}$ and $\hat{y}$ directions respectively, we act with $\oW_{e,+}$; on edges $3$ and $4$, where the arrow points along $-\hat{x}$ $-\hat{y}$, we act with $\oW_{e,-}$.  }  \label{BPFig}
\end{center}                
\end{figure}

\begin{figure}
\begin{center}
\includegraphics[width=.9\linewidth]{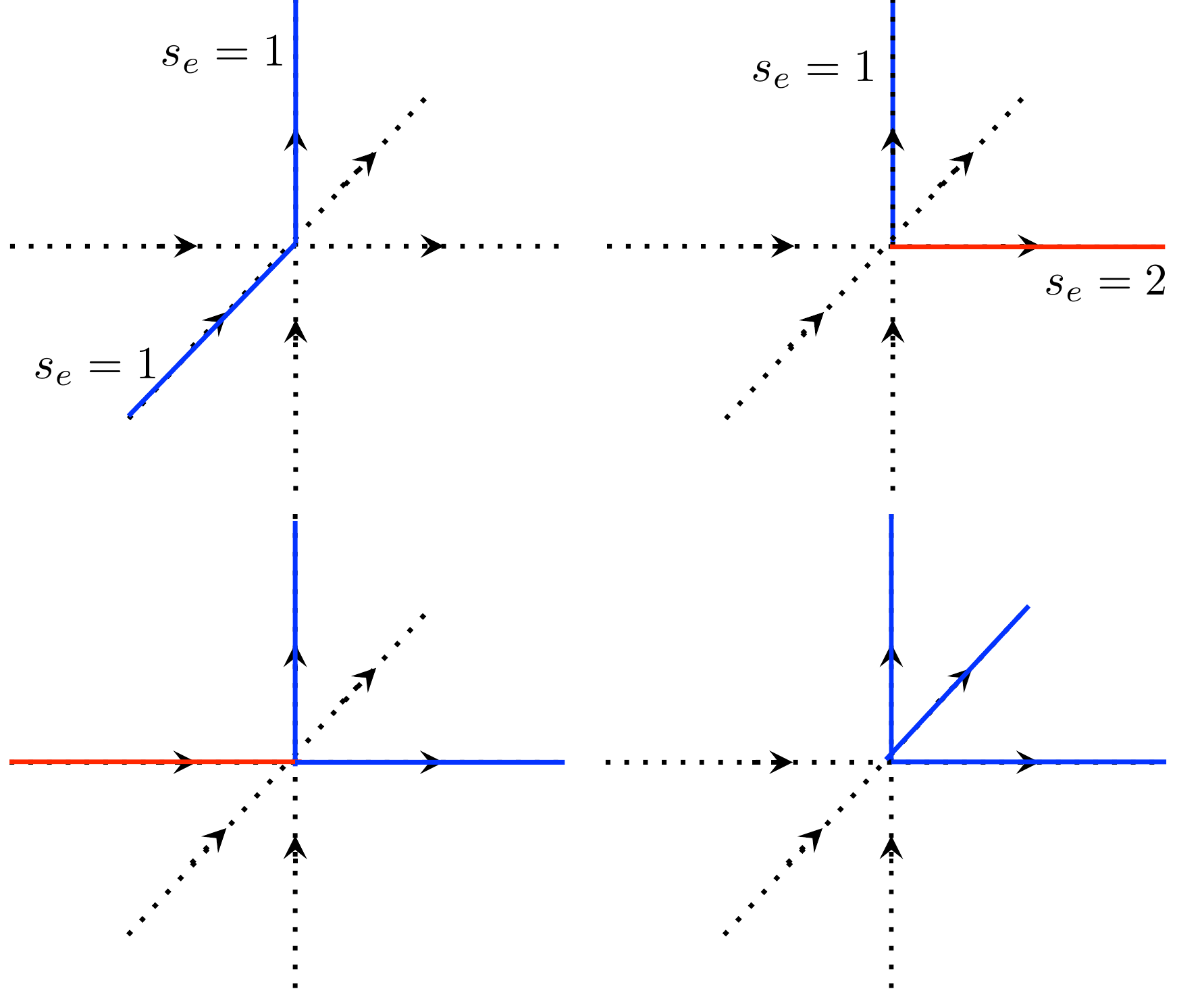}\\
 \caption{(Color online): This figure gives a selection of the configurations satisfying the vertex condition $\sum_{e \in *V} \left( \nu_e s_e \right) =0$ in a $p=3$ Potts gauge theory. Configurations are represented by making edges with $s_e=1$ blue, $s_e=2$ red, and those with $s_e=0$ dotted. The orientation of each edge is indicated by an arrow: on edges where the arrow is entering (leaving) the vertex, we take $\nu = -1$ ($\nu = 1$).  
 }  \label{pottsvertex}
\end{center}                
\end{figure}

The models we describe are a generalization of the 3D Ising gauge theory\cite{WegnerJMP12}, or 3D Toric Code\cite{HammaPRB72}, to Potts spins.  The Hilbert space of our model thus consists of a $p$-state system (a Potts spin $s$) on every edge of a 3D cubic lattice. To define the Hamiltonian we require two types of operators which act on each edge:  $\oS$, which measures the spin, and $\oW$, which raises it. 
We will generally work in the spin basis, where a state in the Hilbert space is specified by the value of the spin on each edge. 
In this basis, on each edge we require the four unitary operators
\ba \label{Sops}
\oS_{e,\pm} |s_{e} \rangle &=& e^{\pm i 2 \pi s_{e}/p }  |s_e \rangle \ \ \ \ \ \ \ \  \ s_e= 0, 1, ... p-1 
\n
\oW_{e, \pm} |s_e \rangle& =&  |s_e\pm 1 \text{ ( mod } p) \rangle
\ea
which obey
\ba \label{HComs}
\oS_{e_1, + } \oW_{e_2, + } &=& \text{Exp}\left[ i 2 \pi/ p  \ \delta_{e_1, e_2} \right ] \oW_{e_2, + } \oS_{e_1, + } \n
\oS_{e_1, - } \oW_{e_2, + } &=& \text{Exp}\left[- i 2 \pi/ p  \ \delta_{e_1, e_2} \right ] \oW_{e_2, + } \oS_{e_1, - } \n
\left( \oS_{e, \nu} \right)^p &=& \left( \oW_{e, \nu} \right)^p =\hat{\bf{1}}
\ea
We require four operators on each edge, rather than two, in order to be able to associate our spin variables with fluxes: in what follows, 
we may view $\oS_+$ as measuring the electric flux in the $\hat{x}, \hat{y}$, or $\hat{z}$ directions, and $\oS_-$ as measuring flux in the $-\hat{x}, -\hat{y}$, or $-\hat{z}$ directions; similarly $\oW_{e,\pm}$ measures $e^{\pm  i \int_{e} \vec{A} \cdot \vec{dl}}$, with $\vec{dl}$ oriented in the $\hat{x}, \hat{y}$, or $\hat{z}$ direction.   

The Hamiltonian we will study has the form: 
\be \label{HPotts}
H =  \lambda \sum_P (1- \hat{B}_P)  -\Gamma \sum_e \hat{\bf{h}}_e - M \sum_V \hat{\mac{Q}}_V 
\ee

The three operators are defined as follows.  
The edge operator $\hat{\bf{h}}$ is:
\be \label{ehat}
\hat{\bf{h}}_e = \frac{1}{p}  \sum_{m=0}^{p-1} \left( \oS_{e,+} \right)^m
 \ee
 $\hat{\bf{h}}_e$ is diagonal in the spin basis, with eigenvalue $1$ if $s_e = 0$, and 0 otherwise.
This term therefore assigns an energy penalty of $\Gamma$ to all edges with non-zero spin.

The operator $\hat{B}_P$ acts on a plaquette $P$ according to
\ba \label{BpPotts}
\hat{B}_P& =& \frac{1}{p} \sum_{m=0}^{p-1} \left ( \hat{\Phi}_P\right )^m \n
\hat{\Phi}_P & =&  
\prod_{e \in \partial P} \oW_{e, \nu_e} 
\ea
Here $\partial P$ is the set of edges in the boundary of $P$, oriented clockwise relative to the $\hat{x}, \hat{y},$ or $\hat{z}$ normal, and  $\nu_e$ is positive for edges in $\partial P$ oriented in the $\hat{x}, \hat{y}$, or $\hat{z}$ directions, and negative for edges oriented in the $-\hat{x}, -\hat{y}$, or $-\hat{z}$ directions (see Fig. \ref{BPFig}).  The relations (\ref{HComs}) imply that $\left( \hat{\Phi}_P\right)^m$ simultaneously raises (lowers) the spin on every edge of $P$ traversed in a positive (negative) direction by $m$ units. It follows that $\left(\hat{\Phi}_P\right)^p =1$, and the flux (or eigenvalue of $\hat{\Phi}_P$) through the plaquette has $p$ distinct values, $e^{ 2 \pi i \phi_P /p}, \ \phi_P= 0... p-1$. Moreover, $\hat{B}_P$ assigns an energy penalty of $1$ to states with nonzero flux $\phi_P \neq 0$, thus favoring plaquettes with zero flux.

Finally, the vertex operator is
\be \label{PottsVert}
\hat{ \mac{Q} }_V \equiv \frac{1}{p} \sum_{m=0}^{p-1} \left(  \prod_{e \in *V } \oS_{e, \nu}  \right)^m
\ee
where $*V$ is the set of all edges entering the vertex $V$.  
Here $\nu$ is negative for edges whose separation from $V$ is in the  $-\hat{x}, -\hat{y}$, or $-\hat{z}$ directions, and positive otherwise.  $\hat{\mac{Q}}$ assigns an energy penalty of $1$ for each vertex at which $\sum_{e \in *V} (\nu _e s_e)  \neq 0$ (mod $p$).   

The Hamiltonian (\ref{HPotts}) has a few key features worth emphasizing.  First, 
\be
\left[  \hat{ \mac{Q} }_V, \hat{B}_P\right ]  = \left[  \hat{ \mac{Q} }_V,\hat{\bf{h}}_e\right ] =\left[  \hat{ \mac{Q} }_V,  \hat{ \mac{Q} }_{V'} \right ]  =0
\ee 
and $\hat{ \mac{Q} }_V$ is conserved separately at each vertex. We may therefore restrict our Hilbert space to states for which $\hat{ \mac{Q} }_V =1$, which is equivalent to studying the pure gauge theory analyzed in Ref.'s \onlinecite{PottsGauge1,PottsGauge2}.  This imposes a constraint on the values of $s_e$ about a vertex, illustrated in Fig. (\ref{pottsvertex}).  For $p=2$, for example, edges with $s_e =1$ must form closed loops.    If we include configurations that violate this constraint, the model (\ref{HPotts}) is equivalent to a gauge theory with Potts-spin matter sources that cost energy $M$ at each vertex.

Second, the Hamiltonian (\ref{HPotts}) has two solvable points.  If we take $\lambda =0$, then the Hamiltonian is diagonal in the spin basis, with $\hat{\bf{h}}_e$ assigning an energy cost to any edge with non-trivial spin.  In this limit the ground state is the product state with all $s_e \equiv 0$; loops with $s_e = m$ are low-lying excitations, whose energy scales linearly with the loop length.

The model can also be solved exactly for $\Gamma =0$, since
\be
\left[  \hat{B}_P, \hat{B}_{P'} \right ]  =\left[  \hat{ \mac{Q} }_V,  \hat{ \mac{Q} }_{V'} \right ]  = 0 
\ee
and the Hamiltonian consists only of commuting operators.  Here the ground state satisfies the condition $\hat{\Phi}_P =0$ on every plaquette; low-lying excitations are plaquettes for which $\phi_P = 1, 2, ... p-1$. 
Because $\hat{B}_P$ raises the edge spins, for $\Gamma =0$ each edge has an equal probability of being in any one of the $p$ different spin eigenstates.  These ground states, which are superpositions over all spin configurations for which $\oQ_V$ has eigenvalue $1$ at every vertex, are one example of what is generally called a {\it loop gas} for $p=2$, or a {\it string-net} for $p>2$.

Between these two solvable points, the model undergoes a single phase transition, in which the non-zero spins are confined by the proliferation of vortex loop defects. To understand the two phases, let us begin with the $\Gamma=0$ exactly solvable point.  Here the ground state $|\Psi_0 \rangle_{\Gamma=0}$ satisfies:
\be
\hat{B}_P |\Psi_0 \rangle_{\Gamma=0} = |\Psi_0 \rangle_{\Gamma=0}   \ \  \text{ i.e. }  \hat{\Phi}_P|\Psi_0 \rangle_{\Gamma=0} = |\Psi_0 \rangle_{\Gamma=0} \punc{.}
\ee
Using the relations in Eq. (\ref{HComs}), it is easy to show that  $\hat{\bf{h}}_e$ and $\hat{B}_P$ fail to commute whenever $e\in \partial P$. Indeed, 
\ba
\hat{\Phi}_P |\Psi \rangle = |\Psi \rangle& \ \ \ \ \ \ \Rightarrow  & \hat{\Phi}_P   \oS_{e,+} |\Psi \rangle  =  e^{- i 2 \pi \nu_{e}/p}  \oS_{e,+} |\Psi \rangle  \punc{.}\n
 \ea
This implies that acting on $|\Psi_0 \rangle_{\Gamma=0}$ with $( \oS_e )^m$ creates a small vortex loop of flux $-2 \pi m \nu_{e}/p$ around the edge $e$, thus exciting all plaquettes bordered by $e$. In the deconfined phase these vortex loops, though present microscopically in the ground state for $\Gamma \neq 0$, remain short.  As $\Gamma/\lambda$ grows, eventually these vortex loops proliferate, and the system undergoes a transition into the confined phase.  

Because the $\integ_{p}^\text{Potts}$ models are self-dual and undergo a single transition as a function of $\Gamma/\lambda,$\cite{PottsGauge1} one can formulate a similar, dual description of the same transition starting from the $\lambda=0$ exactly solvable point.  Here the ground state has $s_e \equiv 0$, satisfying:
\be
\hat{\bf{h}}_e |\Psi_0 \rangle_{\lambda=0} =  |\Psi_0 \rangle_{\lambda=0} \ \  \text{ i.e. } \oS_{e,+}  |\Psi_0 \rangle_{\lambda=0} =  |\Psi_0 \rangle_{\lambda=0}\punc{.}
\ee
 Acting on the ground state for $\lambda =0$ with $(\hat{\Phi}_P)^m $ creates an excited state in which a small  loop of edges bordering the plaquette $P$ carry non-zero Potts spin $m$. In the confined phase, these loops (or nets, for $p>2$) of spins remain small.  As $\Gamma/\lambda$ shrinks, eventually these loops (or nets) proliferate, and the system undergoes a transition into the deconfined phase.    

Thus the phase transition separating the confined and deconfined phases can be viewed either as a proliferation of vortex loops (for $\Gamma/\lambda$ increasing) or a proliferation of non-zero spins (for $\Gamma/\lambda$ decreasing).  
This transition  is an example of a transition that is outside of the scope of the standard Landau-Ginzburg paradigm: there is no local order parameter that takes on a non-zero expectation value as we cross the transition\cite{WegnerJMP12}.  Rather, we may probe the onset of confinement by the (non-local) Wilson loop operator\cite{FradkinShenker}.  
To understand this order parameter, we must consider states in which 
$\hat{\mac{Q}}_{V_i} |\Psi \rangle = 0$ at a pair of vertices $V_1, V_2$. In the gauge theory language, we can think of these states as having a pair of test charges at $V_1$ and $V_2$, which we can create by acting with
\be \label{IAmWilson}
\oW^{(q)}_{1, 2} = \prod_{e_k \in C_{12} } \oW^q_{e_k, \nu_k}\punc{,}
\ee
 where $C_{12}$ is a curve connecting vertices $V_1$ and $V_2$.\footnote{ The operator (\ref{IAmWilson}) is not gauge invariant if $V_1 \neq V_2$.  However, it can be rendered gauge 
 invariant by including matter fields at the vertices, and appending appropriate raising operators for these matter fields to the product.  The operator we use is a gauge-fixed version of this.}  (Here $\nu_k = 1$ if $C_{12}$ crosses the $k^{th}$ edge in the $\hat{x}, \hat{y},$ or $\hat{z}$ direction, and $-1$ otherwise; the $p-1$ possible choices of $q$ correspond to the possible charges $1, ... p-1$ of 
 the discrete gauge theory).  In the limit $\Gamma =0$, the energy for creating a pair of charges is always $2M$, irrespective of their separation; throughout the deconfined phase their energy scales like $|\vec{r}_{12}|^0$ at large separations. Conversely, as $\Gamma/ \lambda \rightarrow \infty$ the 
 energy of separating the charges clearly grows linearly with $|\vec{r}_{12}|$; this scaling holds at large separations throughout the condensed phase.  In other words, the Potts gauge theory has $p$ distinct charges (including $0$), which are deconfined in the uncondensed phase, and confined in the condensed phase.  
The Wilson loop operator diagnoses this change in energy cost. %\cite{WilsonLoop}. 

For our purposes a closely related diagnostic -- the topological orders of the two phases -- will prove more practical.  
For $\lambda =0, \gG>0$, the ground state is the unique spin-polarised state with $s_e =0$, in which $\hat{\bf{h}}_e$ has eigenvalue $1$ on each edge. In contrast, we will soon see that when $\lambda >0, \Gamma =0$ the ground state has a degeneracy of $p^3$ for periodic boundary conditions.  This change in the ground state degeneracy indicates that the uncondensed phase is topologically ordered, while the condensed phase (in which edge spins are confined) is not. Topological order cannot change without a phase transition \cite{Hastings2}, so the system must undergo a phase transition as $\Gamma/\lambda$ increases from $0$ to $\infty$.  
  
The phase diagram of the \zp models has been studied in detail both numerically\cite{PottsGaugeNumerics} and through large $p$ series expansions\cite{PottsGauge1,PottsGauge2}.  These results confirm that there are indeed two phases, each of which is adiabatically connected to one of the solvable points discussed above; for $p\geq 2$, in 3D these are separated by a first-order transition.    

It is worth noting that in 2D these models are dual to the more familiar transverse-field Potts model\cite{WuWang}.  In this dual description the phase with large $\Gamma/\lambda$ is paramagnetic, with spins tending to align with the transverse field.  Large $\Gamma/\lambda$ corresponds to the ferromagnetic phase.  Edges with $s_e \neq 0$ correspond to domain walls in the Potts ferromagnet.  This relationship, and its relevance to the 2D cousins of the topological lattice models we will treat in the next section, are discussed in detail in Ref. \onlinecite{TSBLong}.

\subsection{Topological order in the Potts gauge theory}

	\begin{figure}
\begin{center}
\includegraphics[width=.9\linewidth]{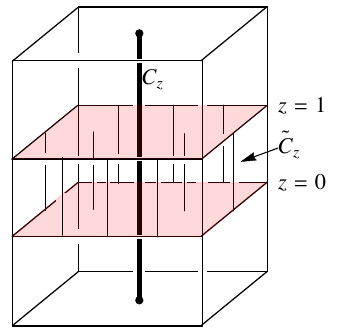}\\
                \caption{(Color online): Ground state degeneracy in the presence of periodic boundary conditions:  With periodic boundary conditions, there exist operators $\oW_{C_\mu}$ that raise spins on all edges along a non-contractible curve $C_\mu$, and commute with the Hamiltonian. (The path $C_z$ is shown in figure.)  Acting with $\oW_{C_\mu}$ changes the ground-state sector.  The operators $\hat{\nu}_{\tilde{C}_\mu}$, which measure all the edges cutting a plane perpendicular to the $\hat{\mu}$ direction, comprise a complete set of labels for the resulting ground states.  In the figure, $\mu=z$ and the edges between $z=0$ and $z=1$ are measured.   }
                \label{sheetop}
\end{center}                
\end{figure}

Hamiltonians exhibiting different topological orders necessarily represent distinct phases of matter\cite{Hastings2}.  Throughout this work we will shed light on the phase diagrams of various systems by distinguishing the topological orders at exactly solvable points in the phase diagrams. Our primary test of topological order is the ground state degeneracy on non-simply connected manifolds, which we now calculate for the Potts gauge theory.

Let begin in a ground state at the exactly solvable point $\Gamma =0$, and consider the effect of the operator
\be \label{WCDef}
\oW_{C_\mu} \equiv \prod_{e \in C_\mu} \oW_{e,+}
\ee
where $C_\mu$ is a closed curve encircling the system in the $\mu =\hat{x},\hat{y},\hat{z}$ direction (see \figref{sheetop}).  
For $\Gamma =0$, $\oW_{C_\mu}$ commutes with the Hamiltonian.  Thus for any ground state, we must have:
\be
 \oW_{C_\mu} |\Psi_0 \rangle = |\Psi_0^{'} \rangle
\ee
with $ |\Psi_0^{'} \rangle$ also a ground state.  We will find that the system is topologically ordered because these ground states are physically distinct.  

To see this, consider the operator
\be \label{nuCEq}
\hat{\nu}_{\tilde{C}_z} =\prod_{e \in \tilde{C}_z } \oS_{e, +}
\ee
where $\tilde{C}_z$ contains all edges connecting the planes $z=0$ and $z=1$.  
$\hat{\nu}_{\tilde{C}_z}$ commutes with $\hat{\Phi}_P$ for every $P$: if the latter raises the spin on one edge in $\tilde{C}$, it simultaneously lowers it on an adjacent edge, preserving the value of the product.  However, 
\be \label{GSEnd}
\hat{\nu}_{\tilde{C}_z} \oW_{C_z} = e^{ i 2 \pi /p } \oW_{C_z} \hat{\nu}_{\tilde{C}_z} 
\ee
as there is exactly on edge on which both operators act.  
The states $\left ( \oW_{C_z} \right)^m |\Psi_0 \rangle, \ m =0, ... p-1$ therefore all have distinct eigenvalues under  $\hat{\nu}_{\tilde{C}_z}$ (and similarly for $x$ and $y$).  We will call the ground state for which $\hat{\nu}_{\tilde{C}_{x}}, \hat{\nu}_{\tilde{C}_{y}}, \hat{\nu}_{\tilde{C}_{z}}$ all have eigenvalue 1 the {\it trivial ground state}.

The topological ground state degeneracy is $n_{\text{free}}^{N_C}$, where $n_{\text{free}}$ is  the number of deconfined charges, and $N_C$ the number of non-contractible curves.  Indeed, the operator $\oW_{C_z}$ defined in Eq. (\ref{WCDef}) can be understood as $\oW^{(1)}_{1 1}$, where $\oW^{(n)}_{ij}$ is given in Eq. (\ref{IAmWilson}), and the curve from $V_1$ to $V_1$ winds around the system in the $z$ direction.  

What is the fate of this ground state degeneracy as the system crosses the phase transition?  
As $\lambda \ra 0$, configurations which have any overlap with the original nontrivial ground states $\hat{\nu}_{\tilde{C}_{\ga} } \neq 1$ (for $\ga=x,y\text{ or }z$) become highly excited states with energy $\Gamma L_z$, which diverges in the thermodynamic limit. This is because any such configuration has a non-contractible cycle (one which wind around the system in the $\hat{x}, \hat{y},$ or $\hat{z}$ direction) on which $s_e \neq 1$. Thus as soon as the phase boundary is crossed, these configurations are eliminated from the low-energy Hilbert space, and the model has no topological order.  

\subsection{Relation to $\mathbb{Z}_p$ lattice gauge theory} \label{ZpClockSect}

It is worth noting that for $p>3$ the operators (\ref{ehat}) and (\ref{BpPotts}) differ slightly from their counterparts in the conventional $\mathbb{Z}_p$ gauge theory, which are given by:
\be \label{ehat2}
\hat{\bf{h}}_e = \frac{1}{2} \left( \oS_{e,+} + \oS_{e, -} \right )
 \ee
\be \label{BpClock}
\hat{B}_P =  \frac{1}{2} \left ( \hat{\Phi}_P +\hat{\Phi}_P^\dag \right ) \n
\ee
where $\hat{\Phi}_P^\dag = \hat{\Phi}_P^{p-1}$.  

Far from the transition, this model has the same physical properties as the $\integ_{p}^\text{Potts}$ model: for $\Gamma=0$, the ground state is flux-free ($\phi_P \equiv 0$), and at small $\Gamma/\lambda$ there is a phase with $p$ deconfined ``charge"-type excitations and a topological ground-state degeneracy of $p^3$ in periodic boundary conditions.  For $\Gamma/\lambda$ very large, similarly, the ground state is a product state $s_e \equiv 0$, in which all spin labels are confined and there is no topological order.  

The main difference between these models and the $\integ_{p}^\text{Potts}$ models we focus on here is at intermediate $\Gamma/\lambda$: for $p \geq 5$, in addition to the two phases of the $\integ_{p}^\text{Potts}$, the \zp gauge theory has a third phase with gapless gauge excitations; this becomes the Coulomb phase of electromagnetism in the limit $p \rightarrow \infty$\cite{ElitzurPRD19,HornPRD19,YukawaGuthPRD21}.  These early analyses further suggest that the two transitions into this gapless phase are second order.\footnote{The order of the transition in the U(1) case has been somewhat controversial, but recent numerical results\cite{CheluvarajaJPA33} support the earlier analytical claim that it is indeed second order.}

\section{Condensation in 3D Confined Abelian Walker-Wang models} \label{AWWSec}

In this section, we will discuss a transition similar to the confining transition of the \zp models of the previous section in a very different family of 3D topological spin models: the Confined Abelian Walker-Wang (\AWW) models discussed in Ref.'s \onlinecite{WW,WWUs}.
Before entering into the details, it is useful to compare \AWW models to those found in the previous section. Like the Potts gauge theories of the previous section, the \AWW models have two phases separated by a first-order transition, which can be viewed as a condensation of vortex loops.  In the uncondensed phase the models have loop gas (for $p=2$) or string-net (for $p>2$) type ground states, with deconfined edge spins; in the condensed phase the non-trivial spins are confined, and (in the limit $\Gamma \rightarrow \infty$) ground states are product states with $s_e =0$ on every edge. 

Unlike in the previous section, we will find that neither phase has conventional topological order. In order to distinguish them using criteria such as Wilson loops or ground-state degeneracy, we will have to study the models in the presence of boundaries.  This is a consequence of the fundamentally different type of topological order found in the Walker-Wang models, which we will call surface topological order.  

\subsection{Walker-Wang Hamiltonians}\label{ss:AWW:Ham}

	\begin{figure}
\begin{center}
\includegraphics[width=.9\linewidth]{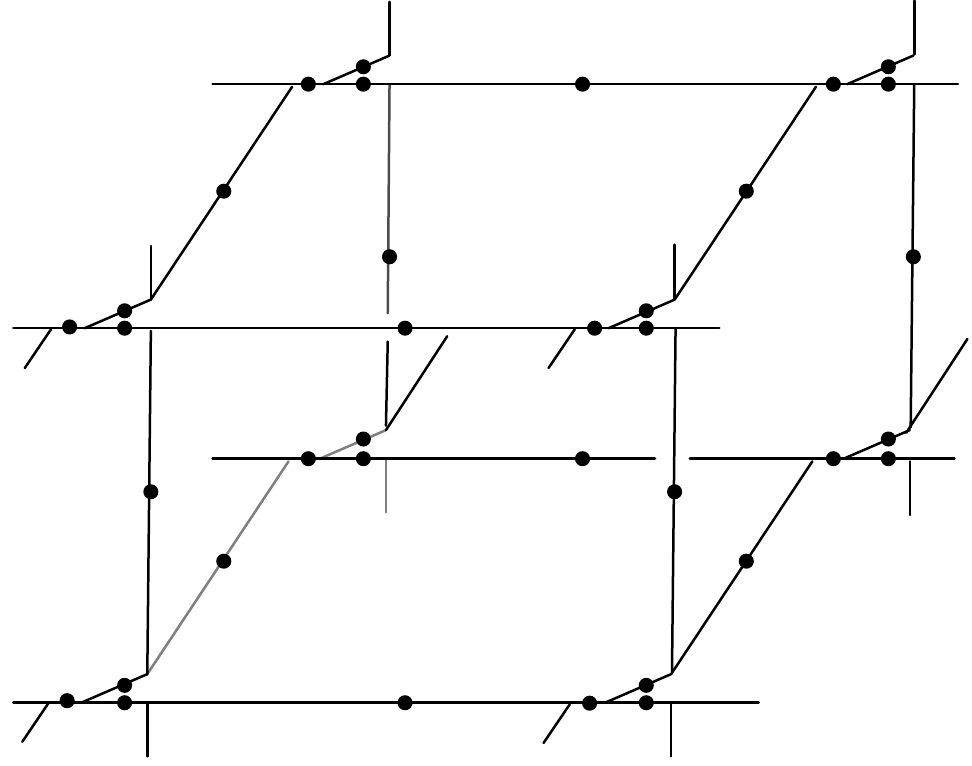}\\
                \caption{ This figure illustrates how the 6-valent vertices on a cubic lattice are split into a collection of trivalent vertices. The spin degrees of freedom living on each edge are represented by black dots.}
                \label{LatticeFig}
\end{center}                
\end{figure}

We begin with a brief review of the Walker-Wang Hamiltonians discussed in depth in Refs \onlinecite{WW,WWUs}.  In this section we will focus on a subset of these, which we call Confined Abelian Walker-Wang (\AWW) models as they are closely related to the Abelian lattice gauge theories discussed in the previous section, but do not have any deconfined bulk excitations (even if we allow vertex defects).   For technical reasons the models are most easily discussed on a lattice for which all of the vertices are trivalent; we thus  deform the cubic lattice by separating each of its hexavalent vertices into three trivalent ones, as shown in Fig. \ref{LatticeFig}.  This deformation has little impact on the physics, so readers may safely imagine that the lattice is cubic unless otherwise notified.

 Like the Potts gauge models of Sect. \ref{PottsSec}, the \AWW models begin with a Potts spin $s = 0,1, ... p$ on each edge of a 3D lattice.  
Schematically, the Hamiltonian has the same form as (\ref{HPotts}): 
\be \label{HWW}
H =  \lambda \sum_P (1- \hat{B}_P)  -\Gamma \sum_e \hat{\bf{h}}_e - M \sum_V \hat{\mac{Q}}_V \punc{,}
\ee
with
\be \label{WWCom1}
\left[ \hat{B}_P, \hat{B}_{p'} \right ] = \left[ \hat{\mac{Q}}_V, \hat{\mac{Q}}_{V'} \right ] =  \left[ \hat{B}_P, \hat{\mac{Q}}_V \right ] =0 \punc{,}
\ee
and
\be\label{WWCom2}
\left[ \hat{\bf{h}}_e, \hat{\bf{h}}_{e'}  \right ]  = \left[\hat{\bf{h}}_e , \hat{\mac{Q}}_V  \right ]  =0\punc{.}
\ee
This guarantees \AWW models share two key features with Potts gauge theories.  First, as $\hat{\mac{Q}}_V $ commutes with all operators in the Hamiltonian, we may consider the limit $M \rightarrow \infty$, in which the constraint $\hat{\mac{Q}}_V =1$ is imposed at each vertex.  Second, the model can be solved exactly both for $\Gamma/ \lambda=0$ and for $\Gamma/\lambda = \infty$.  Using these two limits, we will be able to characterize exactly the topological order of the two phases of this system.

For the \AWW Hamiltonian, we take $\hat{\mac{Q}}_V$ to be given by Eq. (\ref{PottsVert}), and $\hat{\bf{h}}_e$ to be given by Eq. (\ref{ehat}).  
 The only difference between these models and those of Section \ref{PottsSec} is in the definition of the plaquette operator.  For the Walker-Wang models
 \be \label{BpWW}
 \hat{B}_P = \frac{1}{p} \sum_{m=0}^{p-1} \left ( \hat{\Phi}_P\right )^m \hat{\Theta}_{P,m} \punc{,}
\ee
 where $ \hat{\Phi}_P$ is the operator that raises all spins on the edges of $P$, exactly as in Eq. (\ref{BpPotts}).  The difference between $\hat{B}_P$ as given in (\ref{BpPotts}) and (\ref{BpWW}) is the presence of the additional operator $\hat{\Theta}_{P,m}$.  This operator is diagonal in the spin basis, with possible eigenvalues $e^{ i  \pi n/p}, \  n=0, ...2 p-1$; its value depends only the values of the spin variables on certain edges of $P$, and those on the `legs' of the plaquette (i.e. edges that share a vertex with a pair of  edges of $P$). 
   
The precise form of the operator $\hat{\Theta}_{P,m}$ for general $p$ is not central to our discussion, but it is given in Appendix \ref{ThetaApp}.   For our purposes, it suffices to note that, as shown in Appendix \ref{ThetaApp},
 \ba
  \left[ \left( \hat{\Phi}_{P}\right)^m \hat{\Theta}_{P, m} \right]^p& =& 1 \n
 \left[  \hat{\Phi}_{P} \  \hat{\Theta}_{P, 1} \right]^m   &  =&  \left( \hat{\Phi}_{P}\right)^m \hat{\Theta}_{P, m}
  \ea
 The spectrum of $\hat{B}_P$ is therefore identical to that of the plaquette term in the $\mathbb{Z}_p^{\text{Potts}}$ gauge theory:  Since $\left (  \hat{\Phi}_{P} \  \hat{\Theta}_{P, 1} \right)^p =1$, the eigenvalues of $ \left( \hat{\Phi}_{P} \  \hat{\Theta}_{P, 1}  \right)$ are $\lambda = e^{ i \phi_P }, \ \phi_P=2 \pi \ n /p, \  0 \leq n < p$.   Further,
 \be
 \hat{B}_P = \frac{1}{p} \sum_m  \left[  \hat{\Phi}_{P} \  \hat{\Theta}_{P, 1} \right]^m \ \ .
 \ee
  $\hat{B}_P$ therefore has eigenvalue $1$ on states where $\phi_P =0$ (which we can interpret as states with trivial magnetic flux), and eigenvalue $0$ on states with $\phi_P \neq 0$ (which are thus states carrying non-trivial flux), exactly as in the Potts gauge theory.   
 
 \subsubsection{ Example: $p=2$} \label{Sec2p}

	\begin{figure}
\begin{center}
\includegraphics[width=.9\linewidth]{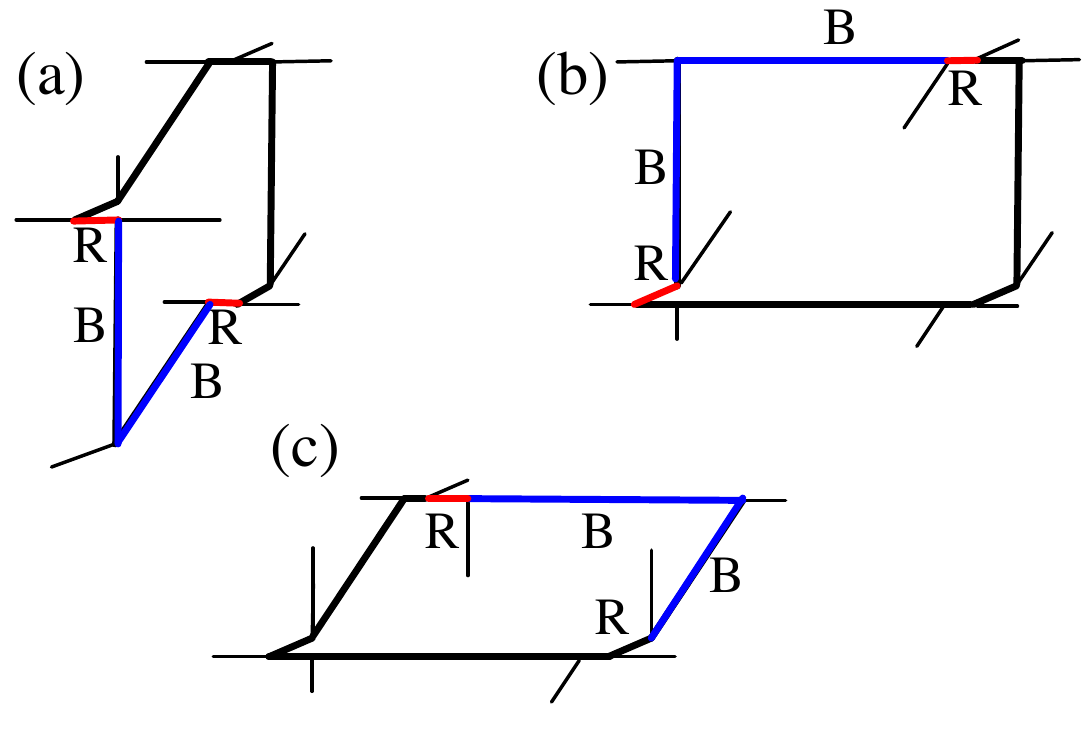}\\
                \caption{(Color online): The plaquette operator for the $p=2$ CAWW Hamiltonian, illustrated for the three different plaquette orientations of the point-split cubic lattice.  The choice of red and blue edges used in Eq. (\ref{BPZ2}) is indicated by the colors (and the letters R and B).  All edges shown in the figure enter into the plaquette operator: edges in the plaquette boundary $\partial P$ are shown in bold; the remaining 10 ``external" edges are in $*P$.  }
                \label{RBFig}
\end{center}                
\end{figure}

For concreteness, consider the example of $p=2$.  This describes a system with an Ising spin $\sigma^z = \pm 1$ on each edge.   We may represent
 \be
  \hat{\mac{Q}}_V  = \prod_{i \in *V} \sigma^z_i \punc{.}
  \ee
 States that satisfy the condition
 \be \label{Qconstr}
 \hat{\mac{Q}}_V  |\Psi \rangle = |\Psi \rangle 
 \ee
 are thus states with down spins (in the $\sigma^z$ basis) on either 0 or 2 of the three edges entering each vertex.   In other words, for $M \rightarrow \infty$, we keep only configurations for which down spins form closed loops on the lattice.  
 
 The plaquette operator for the $\integ_{2}^\text{Potts}$ (or Ising gauge) Hamiltonian is 
 \be
 \hat{B}_P^{\text{Potts}} = \frac{1}{2} \left [ \hat{\bf{1}} +  \prod_{e \in \partial P } \sigma^x_e \right ]
 \ee
 Since the non-identity term acts by simultaneously flipping all of the spins around a plaquette, 
the product over all plaquettes
$
\prod_P \hat{B}_P
$
 contains matrix elements (of equal amplitude) connecting every possible configuration of closed loops with every other.  (Here we are excluding loops encircling non-contractible curves, such as those mentioned in Eq. (\ref{WCDef}), from our definition of ``closed" loops).  
The ground states satisfy
\be
\hat{B}_P |\Psi \rangle = |\Psi \rangle \punc{,}
\ee
and are therefore equal-amplitude superpositions over all possible closed loop configurations on the lattice, with relative coefficient $1$.  
Excited states can be formed by acting with $\oS_{e,+} = \sigma^z_e$ on some number of edges, changing the sign of the coefficients of configurations in which $\sigma^z_e = -1$.  

 The plaquette operator for the Walker-Wang Hamiltonian is\cite{WWUs} 
 \be \label{BPZ2}
 \hat{B}_P^{\text{WW}} = \frac{1}{2} \left [ \hat{\bf{1}} -  \left( \prod_{e \in \partial P } \sigma^x_e \right ) \left(  \prod_{j \in *P } i^{ s_j }  \right ) i^{ s_r+s_{r'} - s_b - s_{b'} }\right ]
 \ee
 where $ *P$ is the set of all ``legs" of  the plaquette $P$, and $r, r'$, $b, b'$ are edges in the boundary of $P$, as shown in Fig. (\ref{RBFig}).   
Here $s_e = \frac{1}{2} \left( 1- \sigma^z_e \right) = 0, 1$ is the value of the Potts spin on edge $e$.  It is not hard to check that $(\hat{B}_P^{\text{WW}} )^2 = \hat{B}_P^{\text{WW}} $, provided that Eq. (\ref{Qconstr}) is satisfied at every vertex. This implies that $\hat{B}_P^{\text{WW}}$ is a projector with eigenvalues $0$ and $1$.

The non-identity component of $ \hat{B}_P^{\text{WW}} $ acts by simultaneously flipping all of the spins around a plaquette, 
and multiplying the resulting configuration by a phase of $\pm 1, \pm i$ which depends on the spin configuration in $*P$ and $\partial P$.  
The ground state is thus a superposition over all possible configurations of closed loops, with coefficients equal in amplitude but with different relative phases $\pm 1, \pm i$.  
As for the $\integ_{2}^\text{Potts}$ model, we can create plaquette excitations by changing the relative phases of these loop configurations.

\subsection{Topological order in the Confined Abelian Walker-Wang models} \label{WWTopSect}

We begin by reviewing the defining features of the \AWW models for $\Gamma=0$.  Since $\hat{B}_P^{\text{WW}}$ commutes with $\oQ_V$, in this limit the models are exactly solvable. The resulting spectrum is discussed rigorously in Ref. \onlinecite{WWUs}; our discussion here is more qualitative, and we refer readers to that work for a detailed treatment.   
 
 The presence of $\hat{\Theta}$ in the plaquette term of the \AWW model (Eq. (\ref{BpWW})) fundamentally changes the nature of the low-lying excitations, as well as the model's topological order.  Unlike the $\integ_{p}^\text{Potts}$ gauge theories, these models have no deconfined excitations in the bulk, even in phases where the spin labels are deconfined and the ground state is a string-net. Commensurate with this, these 3D models have a non-degenerate ground state, even for periodic boundary conditions, as well as a vanishing topological entanglement entropy.
 However, on a lattice with boundaries, the models do have deconfined point particles on the surface and a ground state degeneracy sensitive to the topology of the surface. Hence, the models do exhibit topological order, but it is associated with their surfaces, rather than with the bulk.  
 
 It is worth noting that in CAWW models, generically the surfaces are described by topological orders that can be realized by a purely 2D system, suggesting that the surface topological order can be destroyed without undergoing a bulk phase transition.  In the absence of extra symmetries, therefore, in most cases we know of no criterion which can differentiate the CAWW model from the topologically trivial confined phase.  We will return to this point in Sect. \ref{Conclusion}.  
 
Let us briefly review why the bulk ground states and spectrum differ from those of the $\integ_{p}^\text{Potts}$ models.  
Essentially, the \AWW models have a {\it unique} ground state in the presence of periodic boundary conditions because there is no analogue of the operator
$\hat{\bf{W}}_C$ (Eq. \ref{WCDef}) that commutes with $\hat{B}_{P}^{(\text{WW})}$ for every plaquette $P$ and toggles between the different topological sectors. 

The origin of the differences in the low-lying excitations between the \AWW and $\integ_{p}^\text{Potts}$ models is similar.  For $\Gamma=0$, Eq. (\ref{HWW}) indicates that excitations in the \AWW model can violate either the vertex term (as for the charge excitations in the $\integ_{p}^\text{Potts}$ model), the plaquette term (as for vortex loops in the $\integ_{p}^\text{Potts}$ model), or both.  Plaquette violations behave essentially identically in both models: they must form closed vortex loops, with an energy cost proportional to the loop length. In the \AWW models, however, it is not possible to create a pure vertex violation (see Appendix \ref{ThetaApp}).  Rather, the low-lying bulk excitations in the \AWW models consist of vortex loops (aka closed loops of plaquette defects), and open flux tubes with electric charges at their end-points (literally, a pair of vertex violations separated by a line of plaquette defects).

For instance, in the $p=2$ \AWW discussed above, an operator that raises the spins on two legs of the plaquette $P$ changes the eigenvalue of $\hat{\Theta} \equiv  \left(  \prod_{j \in *P } i^{ s_j }  \right ) i^{ s_r+s_{r'} - s_b - s_{b'} }$ (Eq. \ref{BPZ2}).  (Specifically, it is changed by $-1$ or $\pm i$, depending on whether any of the  $r,r', b$ or $b'$ edges are also  flipped).  Thus any operator that only raises the spins on these edges fails to commute with $\hat{B}_P$.  More generally, we might consider a modified operator, which both raises the spins and multiplies the result by a configuration-dependent phase.  It can be shown\cite{WWUs} that no such operator commutes with $\hat{B}_P$ for all $P$ (with periodic boundary conditions).  This means both that there is no operator $\oW_C$ that commutes with the Hamiltonian, and that vertex defects are confined.  The latter follows because creating a pair of vertex defects at vertices $v_1$ and $v_2$ requires raising the spin by $1$ along a continuous set of edges connecting $v_1$ to $v_2$, and this raising operator necessarily leaves a string of violated plaquettes in its wake.  In other words, the operator that creates a pair of ``charges" (vertex violations) also creates a string of plaquette violations separating these charges, which we interpret as a magnetic flux tube.

The previous discussion is relevant to systems with (for example) periodic boundary conditions.  In systems with open boundaries, the \AWW models have additional excitations that are localized at the surfaces, and a ground-state degeneracy that depends on the surface topology.  A useful intuitive picture of this is that the flux tube connecting the pair of charges can sit just above the surface, where it does not cost energy.  (In technical terms, it is possible to define an operator which creates a pair of vertex defects on the surface without violating any plaquettes.  This operator acts by a combination of raising the spin along a series of edges on the surface, and multiplying by a configuration-dependent phase.)  These surface ``charges" in fact behave like 2D charge-flux bound states, in the sense that they their mutual statistics are those of a $\nu = 1/p$ Laughlin state.  (In more technical terms, these surface states have the topological order of an Abelian Chern-Simons theory).    

Since the charges are deconfined, a process that creates a pair of charges, moves them along the surface, and re-annihilates them does not change the system's energy.  It follows that if there are non-contractible curves {\it on the surface}, there is an operator analogous to $\oW_C$ that raises spins along a closed non-contractible loop and commutes with the Hamiltonian.  In this case, however, such operators do not necessarily commute with each other.  For example,  if we impose periodic boundary conditions in $z$ only, the boundary of the system is a torus with two non-contractible curves, and we may define the operators $\oW_{C_z}, \oW_{C_y}$.  However, these two operators do not commute\cite{WWUs}, and thus are not simultaneously diagonalizable.  The ground state degeneracy is therefore $p$.  Again, this matches the ground state degeneracy of the $\nu=1/p$ Bosonic Laughlin state.  

 \subsection{Condensation transitions in the Confined Abelian Walker-Wang models}

We are now ready to consider the Hamiltonian (\ref{HWW}) for $\Gamma >0$.  In particular, we wish to understand the structure of its phase diagram, and the nature of any phase transitions.  Much of this information will be inferred from our knowledge of the phase diagram of the \zp model (\ref{HPotts}) --despite the significant differences between them for $\Gamma=0$, the phase portraits of the two models are very similar.  

First, note that the effect of the transverse field term $\oh_e$ is the same in both models: it penalizes any edge with $s_e \neq 0$.  In the limit $\Gamma/\lambda \rightarrow \infty$, the ground states of both models are simply the product state $s_e \equiv 0$; for large $\Gamma$ the non-zero spins (and therefore all charges) are confined, and cannot form extended loops or nets.
Thus for sufficiently large $\Gamma/\lambda$ the two models are in the same (topologically trivial) phase.

It is worth pointing out that the mechanism for confinement in the \AWW models with $\Gamma=0$  is fundamentally different from the mechanism for confinement in this trivial phase.  To see this, let us consider the $p=2$ model of Sect. \ref{Sec2p}.  In the trivial phase, the transverse field term $\sum_e \sigma^z_e$ penalizes any edge with $\sigma^z =-1$, meaning that only short loops of these spins can occur.  In the Walker-Wang phase with $\Gamma=0$, conversely, there is no penalty for an edge with $\sigma^z=-1$, and indeed the ground state of this model is a superposition over loops of all sizes.  Hence if we view edges with $\sigma^z =-1$ as occupied by non-trivial electric flux, electric flux is deconfined in the uncondensed phase, and confined in the condensed phase.   However, there are no deconfined point-like excitations in the uncondensed phase, as the model does not admit any excitations that carry only electric charge.  

Second, the transition into this trivial phase is identical in both models.  
As we show in detail in Appendix \ref{PhaseTranApp}, for the edge and plaquette degrees of freedom that are not conserved by (\ref{HWW}), all  correlation functions must be the same.  Essentially, this stems from the fact that the commutation relations between the non-commuting operators in (\ref{HPotts}) and (\ref{HWW}) are identical.  Specifically, for $e \in \partial P$, 
 \be \label{TheZLikeCommutator}
 \left( \hat{\Phi}_P \hat{\Theta}_{P,1}  \right )\oS_{e, \nu_e} = e^{ - 2 \pi i /p } \oS_{e, \nu_e} \left( \hat{\Phi}_P \hat{\Theta}_{P,1}  \right )  \ \ \ 
 \ee
so that in both models $\left(  \oS_e \right )^n$ raises the flux eigenvalue of $\left( \hat{\Phi}_P \hat{\Theta}_{P,1}  \right )$ by $ 2 \pi n /p$, creating one of   $n$ distinct types of vortex loop excitations.     
This can be used to show that the amplitude for finding a given configuration of vortex loops in the ground state at a particular value of $\Gamma/ \lambda$ is thus equal in both models (see Appendix \ref{PhaseTranApp}), which ensures that the transitions are the same.

 An alternative way to understand this result is the following.   The only difference between the two models is the presence of $\hat{\Theta}_{P,m}$ in the plaquette projector, which changes the relative phases of different spin configurations in the ground state.  However, these phases do not affect the expectation values of operators diagonal in the spin basis. (We take different spin configurations to be orthogonal).    Rather, they are only sensitive to the probability that $s_e$ takes on some particular value-- which is the same in both models (provided we work in the trivial ground state sector). 
Put another way, the physical difference between the two models is in the charge sector, with charges confined to the ends of flux tubes in the \AWW case.  The transition, on the other hand, is in the purely magnetic sector, involving a condensation of vortex loops.  From this perspective, it is unsurprising that the nature of the phase transition is the same in both models.

As in the previous section, for $p>3$ we can contemplate a variant of our Hamiltonian, with $\oh_e$ replaced by (\ref{ehat2}), and the plaquette operator given by $\hat{\Phi}_P \hat{\Theta}_{P,1} + h.c.$.  Because the arguments given above apply equally well here (see Appendix \ref{PhaseTranApp}), the thermodynamic properties of the two models are the same.  Hence  for $p \geq 5$ in the ``clock" variant of Eq. (\ref{HWW}), the two confined phases are separated by a critical, Coulomb-like phase.  In this phase neither the vortex loops nor loops of non-trivial spin need have a finite energy cost -- suggesting that the charge (or vertex-violating) excitations also become deconfined in this r{\'e}gime.

\bigskip
 
Before discussing more general transitions in these systems, it is worth pausing to consider what the results of this section tell us about the phase portrait of the Hamiltonian (\ref{HWW}).  In the Potts gauge model, the phase transition separates systems with different topological orders, which are therefore necessarily distinct phases.  In the Walker-Wang system, however, the phase transition connects two systems with trivial bulk topological order.  This raises the question of whether the confined and deconfined ``phases" in the \AWW are truly distinct -- or whether, like the liquid-gas transition, this first-order transition is an artifact of the particular trajectory that our models take through the phase diagram.  

For concreteness, let us focus on the case $p=2$.   In addition to topological order, we might identify two distinct phases by identifying a symmetry that is broken by the \AWW phase but not by the trivial phase, or vice versa.  However, this model possesses only lattice symmetries and the discrete symmetry $PT$, which sends $x \rightarrow -x, y \rightarrow -y, z\rightarrow -z$ and complex conjugates the Hamiltonian.  Because of the nature of the plaquette operator, both $PT$ and the lattice symmetries are symmetries of the system for all values of $\Gamma, \lambda$, while $P$ and $T$ are individually broken unless $\lambda =0$.   

Thus for $p=2$ the \AWW model cannot be distinguished from the trivial confined phase either on the basis of symmetry or on the basis of topological order.  This strongly suggests that there is a deformation of (\ref{HWW}) allowing us to connect the two limits $\Gamma =0$ and $\lambda=0$ without undergoing a phase transition. The arguments in Appendix \ref{PhaseTranApp} ensure that such a deformation must include terms that create vertex violations, as otherwise the free energy is identical to that of a model whose plaquette term contains only the operators $\hat{\Phi}_P^n$, in which the two limits have different topological orders.  
Further, as the phase transition is first order, it will persist for arbitrary small deformations away from (\ref{HWW}) -- possibly ending at a second-order point at some finite value of the perturbation strength.

\subsection{Diagnosing vortex condensation}
How do we discern the behaviour of the \AWW ($\Gamma=0$) from that of the condensed phase ($\Gamma=\infty$)? The condensation transition does not appear to involve the symmetry
breaking of an order parameter and, while the presence of bulk topological
order distinguished the Potts gauge theories from the trivial paramagnet,
we cannot say the same of \AWW models which have no bulk topological
order. One obvious difference between the phases is in the presence
of many vortices, which can be detected by the operator

\begin{equation}
\mac{O}_{\mac{S}}=\prod_{P\in\mac{S}}\hat{\Phi}_{P}\hat{\Theta}_{P,1}
\end{equation}
$\mac{O}_{\mac{S}}$ is formed by acting with the plaquette operators
on open surface $\mac{S}$, with area $A$ and perimeter $L$.
Near the $\Gamma=0$ exactly solvable point, the operator obeys a
perimeter law $\left\langle \mac{O}_{\mac{S}}\right\rangle \sim e^{-\gamma L}$
because vortex loops are short, so can only intersect the surface
if they lie near its boundary. On the other hand, near the $\Gamma\rightarrow\infty$
point, $\left\langle \mac{O}_{\mac{S}}\right\rangle \sim e^{-\sigma A}$,
because vortex loops are larger and proliferate freely. While the
change from area to perimeter law does not imply that the $\Gamma=0$
and $\Gamma=\infty$ points constitute distinct phases of matter,
it does allow us to quantitatively distinguish the two regimes. 

In the Potts gauge theory the fundamental Wilson loop along $\partial\mac{S}$,
which can be written as $\mac{O}_{\mac{S}}^{\mbox{Potts}}=\prod_{P\in\mac{S}}\hat{\Phi}_{P}$,
plays the same role as $\mac{O}_{\mac{S}}$ in tracking vortex condensation.
However, if used in the CAWW case, this operator will exhibit area
law behaviour in both the un-condensed and condensed regimes. This
is because, even at $\Gamma=0$, the magnetic flux measurement $\hat{\Phi}_{P}$
fluctuates wildly -- a defining feature of CAWW models is that only
the special combination $\hat{\Phi}_{P}\hat{\Theta}_{P,1}$ of electric
and magnetic flux measurements is stationary on the ground state.

 \section{Other transitions in Abelian models}\label{s:OtherAbelianTrans}
Thus far, we have identified a set of first order transitions in the $\integ_{p}^\text{Potts}$ and \AWW models which connect a given 3D string-net state to a trivial phase, in which all spin labels are confined.  We now turn to a rather more interesting situation, in which only {\it some} of the spin labels are confined in the condensed phase.  Here we will find a surprising difference between the $\integ_{p}^\text{Potts}$ and \AWW models.  In the $\integ_{p}^\text{Potts}$ case, a transition that confines spin labels  is a confining transition: the number of deconfined charges in the gauge theory decreases, as does the ground-state degeneracy; at long wavelengths the confined phase may be described by an Abelian lattice gauge theory with a smaller gauge group than the deconfined phase.  For the \AWW models, however, in the uncondensed phase there are no deconfined charges, and we might guess that the transitions are necessarily between two phases with trivial bulk topological order and no deconfined point-like excitations.  We will find -- somewhat counter-intuitively -- that this guess is incorrect: a phase transition which confines a subset of the spin labels in an \AWW model can lead to deconfinement in the charge sector.  Correspondingly, we will find that the ground-state degeneracy {\it grows} as we cross the transition into the confined phase.  We will give a physical interpretation of this in Sect. \ref{GenAWWSect}. 

How do we obtain transitions that confine only some of the labels?  
The key is that Eq. (\ref{ehat}) is not the only possible choice of $\hat{\bf{h}}_e$.  More generally, if $p$ is divisible by $m$ we can take
\be \label{hm}
\hat{\bf{h}}_e^{(m)} = \frac{m}{p}  \sum_{n=0}^{p/m-1} \left( \oS_{e,+} \right)^{n m}
 \ee
 which has the effect of proliferating vortices with fluxes $2\pi /p \times \{m,2m,\ldots, p-m \}$. The operator $\hat{\bf{h}}_e^{(m)} $ has eigenvalue $1$ on edges where $s_e$ is a multiple of $p/m$, and 0 otherwise, and therefore assigns an energy penalty $\Gamma$ for every spin that is not a multiple of $p/m$. Thus even as $\Gamma \rightarrow \infty$, some spins remain deconfined.  The result for the $\integ_{p}^\text{Potts}$ models is a condensed phase that has the bulk topological order of a $\mathbb{Z}_{p/m}$ gauge theory.  For the \AWW models, depending on the values of $p$ and $m$, we find that the condensed phase can have bulk topological order, surface topological order, or both.  A table classifying the topological orders of these phases can be found at the end of Sect. \ref{GenAWWSect}.
 
As in the previous sections, ``clock"-like versions of the Hamiltonians that we discuss also exist, and for $m \geq 5$ these will exhibit gapless phases at intermediate values of $\Gamma/\lambda$.  Though this possibility is certainly intriguing, our focus here will be on the possible phases at large $\Gamma$, where the two models are in the same phase.  
 
For readers familiar with TQFTs, we note that not all Abelian Walker-Wang models are completely confining at $\Gamma=0$.  In Appendix \ref{complicatedWWmodels}, we classify the possible condensed phases in arbitrary Abelian Walker-Wang models.  Like the \AWW models discussed here, the structure of their phase diagrams is identical to that of an appropriate $\mathbb{Z}_p^{\text{Potts}}$ gauge theory.
 
 \subsection{Variants on the 3D Potts gauge transition} \label{PottsTranVar}

For pedagogical reasons, we will first describe these transitions in the $\integ_{p}^\text{Potts}$ models. If we replace $\hat{\bf{h}}_e$ by $\hat{\bf{h}}_e^{(m)}$ in Eq. (\ref{HPotts}) and study the resulting phase diagram, we again find two phases separated by a first-order phase transition.  We will show momentarily that an effective Hamiltonian describing the transition can be mapped onto that of a system with $p/m$ spin states and a transverse field $\hat{\bf{h}}_e^{(1)}$, such that these transitions are always first order.   However, the topological order of the phase at large $\Gamma/ \lambda$ is different for every $m$: it is that of an $m$-state Potts gauge theory.  Thus with a Hamiltonian of the form (\ref{HPotts}) and the modified definition of $\hat{\bf{h}}_e^{(m)}$, we can describe a family of transitions between our $p$-state Potts gauge theory and an $m$-state Potts gauge theory, for any $m$ that divides $p$.

We now explain why the phase transition is still of the type described in the previous section.  Since 
\be
\left( \oS_{e,+} \right)^m \hat{\Phi}_P  = e^{ i 2 \pi m/p} \hat{\Phi}_P\left( \oS_{e,+} \right)^m 
\ee
we have
\be \label{HCom2}
\left[ \hat{\bf{h}}_e^{(m)}, \left( \hat{\Phi}_P \right )^{p/m} \right ] =0 \ \ \ \ \ \ \ \ \ \ 
\ee
Because $\left( \hat{\Phi}_P \right )^{p/m}$ commutes with all other terms in the Hamiltonian, 
we may restrict our attention to states $|\Psi \rangle$ for which 
\be \label{mCond}
\frac{1}{m}  \sum_{n=0}^{m-1} \left ( \hat{\Phi}_P\right )^{ n p/m}  |\Psi \rangle =  |\Psi \rangle
 \ee
To ensure that this condition is satisfied by all ground states even for $\lambda=0$, it is convenient to modify the plaquette projector somewhat:
\be \label{BLambda}
\lambda \hat{B}_P = \frac{m}{p}\left [ \Lambda+   \lambda \sum_{i=0}^{p/m-1} \left( \hat{\Phi}_P \right)^i \right]  \left (  \frac{1}{m} \sum_{n=0}^{m-1} \left ( \hat{\Phi}_P\right )^{ n p/m}   \right ) 
\ee
The term proportional to $\Lambda$ commutes with all operators in the Hamiltonian; thus we may consider the limit $\Lambda\rightarrow \infty$, where the condition (\ref{mCond}) is always satisfied.  \footnote{In fact, for $T=0$, this condition is always satisfied, since the vertex term commutes with all operators in the Hamiltonian. }   
On states satisfying Eq. (\ref{mCond}), we have:
\be
\hat{B}_P |\Psi\rangle = \frac{m}{p} \sum_{i=0}^{p/m-1} \left( \hat{\Phi}_P \right)^i  |\Psi \rangle
\ee
Within this sub-space, then, the Hamiltonian for the transition is equivalent to a Hamiltonian of the form (\ref{HPotts}) for a $p/m$ state spin.  

Thus the energetics of the transition are clearly those of a transition with $p/m$-state spins and $\hat{\bf{h}}_e^{(1)}$.  The only remaining subtlety is whether the number of states at each energy is the same in both systems.  We show in Appendix \ref{CountApp} that though the total number of spin configurations at any energy is of course larger in the $p$-state system, this multiplicity merely alters the normalisation of the ground-state wave function, and plays no role in the thermodynamics of the phase transition (at any temperature).  

Next, let us verify that the topological order of the phase for large $\Gamma/\lambda$ is indeed that of an $m$-state Potts gauge theory.  From Eq. (\ref{HCom2}), we see that spins with $s_e = n \ p/m$ carry no energy cost.  It follows, from arguments similar to those given in Eq's (\ref{WCDef}) to (\ref{GSEnd}), that the system has an $m^3$ ground-state degeneracy in periodic boundary conditions\footnote{The ground state degeneracy does not uniquely specify the topological order, but it is straightforward to verify that the statistics of the charge and vortex excitations of the model also match}.  
Similarly there are $m$ deconfined charges in the system, which lie at the end-points of string operators corresponding to the $m$ deconfined spins.

 \subsection{Variants on the transition in Abelian Walker-Wang models} \label{GenAWWSect}
 
We now consider transitions in the \AWW models where the edge term is given by  $\oh^{(m)}$ (Eq. (\ref{hm}) ), for $m$ a divisor of $p$.  
Again, this choice assigns no energy cost to the $p/m$ spins $s = 0, m, 2 m, ... p-m$, while all other spins are eliminated from the low-energy Hilbert space as $\Gamma/\lambda \rightarrow \infty$.  Since $\hat{\bf{h}}^{(m)}$ is insensitive to the relative phases in the ground state wave functions, the  arguments of Sect. \ref{PottsTranVar} and Appendix \ref{PhaseTranApp} can be combined to show that the phase transition is equivalent to that of a $p/m$ state Potts gauge theory restricted to its trivial ground-state sector -- and therefore must be first order.

Our main interest in this section, then, will be to describe the topological order of the condensed phase.
Since the topological order is a property of the phase, it is convenient to study it by identifying a point in this phase at which the Hamiltonian is exactly solvable.  To do so, as in the $\integ_{p}^\text{Potts}$ case  (Eq. \ref{BLambda}).
 for $m>1$ we will modify the Walker-Wang plaquette projector according to:
\ba \label{BpWW2}
\lambda \hat{B}_P& =& \frac{1}{p}\left [ \Lambda  \sum_{n=0}^{m-1} \left ( \hat{\Phi}_P\right )^{ n p/m}   \hat{\Theta}_{P, np/m}  \right . \n
&& \left. +  \lambda \sum_{i=1}^{p/m-1}  \sum_{n=0}^{m-1} \left( \hat{\Phi}_P \right)^{i+ n p/m}  \hat{\Theta}_{P, i + n p/m}\right]  
\ea
. The term proportional to $\Lambda$ commutes with $\hat{\bf{h}}^{(m)}$ irrespective of the presence of $\hat{\Theta}$: $\hat{\Theta}$ and $\hat{\bf{h}}^{(m)}$ necessarily commute, as both are diagonal in the spin basis, and also $[ \hat{\bf{h}}^{(m)},  \left ( \hat{\Phi}_P\right )^{ n p/m} ]=0 $.  Thus the model can be solved exactly in the limit $\lambda =0, \Gamma \rightarrow \infty, \Lambda>0$.
 
For $\lambda =0$, the plaquette projector changes the values of the spins only by multiples of $p/m$.  Taking $\Gamma \rightarrow \infty$ ensures that the low-energy Hilbert space contains only the $m$ spins $s=0, p/m, ... p-p/m$.   Thus by re-labelling the spins $0\rightarrow 0, p/m \rightarrow 1, ... p/m-1 \rightarrow m-1$, we may express the Walker-Wang Hamiltonian in this limit as
\be \label{HWW2}
H_{\text{WW} } = - M \sum_V \hat{\mac{Q}}_V - \Lambda \sum_P \left( 1 -\sum_{n=0}^{m-1} \left ( \hat{\Phi}_P\right )^{ n}   \hat{\tilde{ \Theta}}_{P, n} \right )
\ee
with $\hat{\mac{Q}}_V$, $\hat{\Phi}_P$ the vertex and flux operators of the $m$-state Potts gauge theory, and $\hat{\tilde{ \Theta}}_{P, n}$ defined by the action of the operator $ \hat{\Theta}_{P, np/m}$ on the reduced Hilbert space containing only spins that are multiples of $p/m$.    
 
To proceed further, we must understand the basic properties of $\hat{\tilde{ \Theta}}_{P, n}$.  In particular, we wish to determine whether there is an operator analogous to $\oW_C$ (Eq. (\ref{WCDef}) ) that commutes with the Hamiltonian (\ref{HWW2}).   If so, the system will have multiple ground states, distinguished by their different eigenvalues under the action of the operator $\hat{\nu}_C$ in Eq. (\ref{nuCEq}) (which does commute with $H_{\text{WW}}$).  If not, as shown in Ref. \onlinecite{WWUs}, the ground state is unique. 

The eigenvalues of  $\hat{\tilde{ \Theta}}_{P, n}$ can be deduced from those of  $\hat{\Theta}_{P, np/m}$, 
which are phases depending on the spin configuration of the  external edges of $P$, as well as the spin configurations of certain edges on $P$.   
In the low-energy Hilbert space containing only spins that are multiples of $p/m$, the possible eigenvalues of $\hat{\Theta}_{P, n p/m}$ are $e^{i  \pi  (r / m) (p/m)}$, where $r = 0, 1, ... m-1$.  

There are four cases to consider here.  First, if $p/m^2 \in 2 \mathbb{Z}$, then $e^{i \pi  (r / m )(p/m)} \equiv  1$ for all $r$.  In this case, one can show that $\hat{\tilde{\Theta}}$ acts as the identity operator on all states obeying $\hat{\mac{Q}}_V |\Psi \rangle = |\Psi \rangle$.  If $p / m^2$ is an even integer, the Hamiltonian (\ref{HWW2}) therefore describes an $m$-state Potts gauge theory, with a topological ground state degeneracy of $m^3$ in periodic boundary conditions.

Second, if $p/m^2 \in 2 \mathbb{Z} + 1$, then $e^{i \pi  (r / m )(p/m)} \equiv (- 1)^r$.  
In this case, we can still define an operator $\oW_C$ for each non-contractible $C$ that commutes with the Hamiltonan for any value of $\lambda$, such that the ground-state degeneracy remains $m^3$ in periodic boundary conditions (see Appendix \ref{ThetaApp} for a proof of this).  In fact, we show in Appendix \ref{FermiApp} that for $M< \infty$ this model describes an $m$-state Potts gauge theory coupled to fermionic matter sources.  

Third, if $m$ is not a factor of $p/m$, then the only value of $r \in \{ 0, ... m-1 \}$ for which $\hat{\Theta}_{P, n p/m}$ has eigenvalue $1$ is $r=0$.  As we show in Appendix \ref{ThetaApp}, this means that the Hamiltonian (\ref{HWW2}) describes an $m$-state \AWW model, with trivial topological order.  

Finally, suppose $p/m^2 = a/b$, with $a$ and $b$ relatively prime, and $b>1$ a factor of $m$.  In this case $\hat{\tilde{\Theta}}$ has $b$ eigenvalues $\pm 1$ for $ r = t b, \ \ t = 0, 1, ... m/b -1$.  If $\hat{\Theta}$ has eigenvalue $1$ for $r \neq 0$, the eigenvalue of $\hat{\tilde{\Theta}}$ is unaffected by replacing $s \rightarrow s + r$ on any edge (see Appendix \ref{ThetaApp}).   Therefore the operator $\left( \hat{W}_C\right)^r$ commutes with the Hamiltonian.  Less trivially, one can show that provided $\hat{\mac{Q}}_V |\Psi \rangle = |\Psi \rangle$ (which is true in the ground states, provided $M$ is positive), $\left( \hat{W}_C\right)^r$ also commutes with $H$  for $r$ corresponding to negative eigenvalues of $\hat{\tilde{\Theta}}$.  
Thus in this case we find a model whose Hilbert space is described by $m$-state Potts spins on each edge, with (in periodic boundary conditions) a ground-state degeneracy of $(m/b)^3$, where $b$ divides $m$.  The topological order of this phase is therefore different from that of both the $m$-state Potts gauge and \AWW models.  

A summary of these possibilities, together with the topological order of the condensed phases, is given in Table \ref{ZpTable}.

\begin{table}[ht]
\begin{tabular}{|c|c|}
\hline
$p/m^2$  &  Topological order of the condensed phase\\
\hline
$2 n,\  n \in \mathbb{Z}$ & $\mathbb{Z}_m$ gauge theory \\
$2 n+1,\  n \in \mathbb{Z}$ &  fermionic $\mathbb{Z}_m$ gauge theory\\
$n/m$ (irreducible) &  $\mathbb{Z}_m$ \AWW (modular)\\
$n/b, \ b| m$ &   $\mathbb{Z}^{m n/(2 b)}_m$ \AWW (non-modular)\\
\hline
\end{tabular}
\caption{\label{ZpTable} Topological orders of $p$ state \AWW models in the condensed phase with transverse-field $\oh^{(m)}$.  Our notation conventions are drawn from Ref. \onlinecite{BondersonThesis}.   } 
\label{tab:CAWW}
\end{table}

What becomes of the excitations as we cross the boundary from an uncondensed phase to a condensed phase?   Since all spins are now multiples of $p/m$, there are only $m-1$ physically distinct vortex-loop creation operators 
\be
\oh^{(1)}, ... , \ \oh^{ p/m -1} \ \ \ \ \ .
\ee
Writing the plaquette projector as in Eq. (\ref{BpWW2}) allows us to separate the fluxes bound to our charges into two kinds: those whose energy cost remains finite (proportional to $\Lambda$) as $\lambda \rightarrow 0$ and those whose energy cost vanishes for $\lambda =0$.  The number of fluxes with zero energy cost in the confined phase is given by the number of spins $r$ for which $e^{ i \pi r\  p/m^2} = \pm 1$.  
Thus some particles become deconfined in the condensed phase, because they lie at the end-points of flux tubes whose energy cost vanishes as $\lambda / \Gamma \rightarrow 0$.   Physically, this is because these flux tubes become unobservable (and thus have vanishing energy) once a subset of the spins have been eliminated from the theory.   The deconfined particles are always fermions or bosons: the criteria $e^{ i \pi r\  p/m^2} = \pm 1$ ensures that the relevant charge-creation operators either commute or anti-commute.  (For a more detailed discussion of the fermionic case, see Appendix \ref{FermiApp}).

It is also instructive to consider the fate of the deconfined surface excitations in the condensed phase.  There are three possibilities: a given type of surface anyon may become confined; may remain deconfined at the surface only; or may become deconfined in the bulk (as well as at the surface).  

To understand the first possibility, we need only know that the surface anyons exist at the end-points of a string of edges of appropriate spins.   For example, in the $p=2$ example discussed above, a pair of surface anyons occurs at the end-points of a string of edges on which $s=1$.  In that case, in the condensed phase edges with $s=1$ bear an energetic cost; hence the surface anyons are confined (in the same sense that charges become confined in the $\integ_{p}^\text{Potts}$  gauge theory), and the transition is one to a state with trivial bulk and trivial surface topological order.  

If the anyon is to sit at the end-point of a string of edges with spin labels that remain deconfined, then that excitations of this type are deconfined on the surface in the condensed phase.  Whether they are also deconfined in the bulk then depends on whether  $p/m^2$ is an integer.  If it is, then our ``anyons" are in fact bosons or fermions, and in the condensed phase have trivial braiding with all of the deconfined surface anyons.  If $p/m^2$ is not an integer, then the condensed phase has excitations that are deconfined only at the boundary of the 3D lattice; at least one of these must be anyonic.

\subsection{Examples}

For concreteness, we will now consider one example from each of the four classes in Table (\ref{ZpTable}).  
To obtain a physical picture of which excitations are deconfined in the bulk, it is helpful to classify the excitations in the Walker-Wang models in terms of their ``charge" (given by the eigenvalue of $\oQ_V$ at the violated vertices) and ``flux" (given by the phase of the eigenvalue of $\hat{\Phi}_{P,1} \hat{\Theta}_{P,1}$ on the line of excited plaquettes connecting the two vertices).  In the uncondensed phase, a charge $q$ lies at the end of a flux tube with flux $\phi = 2 \pi p q$.  The operator that raises edge spins of $P$ by $s$
essentially measures the Berry phase of an object of charge $q = s/p$ around the flux through the plaquette $P$, penalizing states where this Berry phase is not a multiple of $2 \pi$.  To understand why some charges become deconfined in the condensed phase, we observe that for $\lambda =0$, the plaquette term (\ref{BpWW2}) contains only a subset of such possible measurements -- those for charges $q   = n p/m, n=1, ... m-1$.  In this limit, some flux tubes become physically unobservable: there are no longer any charges in the theory with which they have non-trivial Berry phase.  This gives us a simple mnemonic for understanding the topological order of the phase with large $\Gamma/\lambda$.

Armed with this intuition, we will consider each of the four classes (see Table \ref{tab:CAWW}) in turn.  For simplicity, we will discuss these in the solvable limit $\lambda =0, \Gamma>0$, for which there is zero amplitude to create any edges with confined spin labels; however, the physical properties are not restricted to this special point but characterize the condensed phase.

The first possibility is that the condensed phase can be described by a $\mathbb{Z}_m$ Potts gauge theory.  To see this, consider the example $p=8, m=2$.  
We can view the spin labels $s =0, ... 7$ as representing $7$ possible electric fluxes that can exist on each edge of the lattice, associated with fractional charges $q_s=0, 1/8, ... 7/8$.  The possible magnetic fluxes in the theory are then $ 2 \pi n, n=1, .. 7$.  
The fundamental excitations in the uncondensed phase are closed flux tubes of flux $2 \pi n$, and open flux tubes with flux $2 \pi (8q ) $ which terminate at vertices with charge $q$.

For $\lambda/\Gamma =0$, only the two spins $s = 0, 4$ remain in the low-energy Hilbert space.  If we keep only these two spins, the vertex condition requires that the number of edges with $s=4$ at each vertex must be even -- in other words, it reduces exactly to the vertex condition of the $\mathbb{Z}_2$ model.  
Further, if we keep only edge spins $s=0, 4$, then $\hat{\Theta}_4 \equiv \mathbf{1}$, and the plaquette projector simply flips all spins around a plaquette.  Thus our solvable Hamiltonian for the condensed phase is exactly that of $\mathbb{Z}_2^{\text{Potts}}$ gauge theory.  
 
What has become of our charge-flux-tube bound states?  In the condensed phase, the only deconfined charge is $q=1/2$, which lies at the end of a flux tube of flux $8 \pi$.  In the uncondensed phase, this magnetic flux tube is physically observable (and indeed costs energy per unit length), since it is measured by lines of electric flux corresponding to charge $q = 1/8, 3/8, ...$.  Electric flux lines corresponding to $q =1/2$, however, cannot distinguish between a flux $8 \pi$ and a flux $0$, and the $q=1/2$ charges are the only deconfined in the condensed phase.  
The  only other low-energy exctation in this phase is the closed vortex loop of flux $2 (2n +1) \pi$, about which a charge $1/2$ has a Berry phase of $\pi$.  (All such vortex loops are physically indistinguishable in the condensed phase).  
This is exactly the spectrum of  $\mathbb{Z}_2$ gauge theory. 

Next,  consider $p=4, m=2$.  The associated electric charges are $q = 1/4, 1/2, 3/4$, bound to the ends of flux tubes of flux $ 2 \pi (4q)$.  For $\lambda/\Gamma =0$, the only remaining electric flux is $1/2$, and the possible charged excitations are $q= 1/2$, bound to flux tubes $4 \pi$.  These fluxes are physically undetectable by objects of charge $1/2$, so that the charges are deconfined.  The condensed phase also has vortex loop excitations, of flux $2 \pi \equiv 6 \pi$.  
Indeed the only difference from the previous example is that the charge $1/2$ has Berry phase $2 \pi$, rather than $4 \pi$, with its associated flux tube.  As we show in Appendix \ref{FermiApp}, this describes a $\mathbb{Z}_2$ gauge theory with fermionic sources.

 Third, consider $p=6, m=2$.  Here again the electric flux that remains for $\lambda / \Gamma =0$ is $1/2$; charge $1/2$ excitations are now bound to the end-points of tubes of flux $6 \pi$.  These flux tubes have a Berry phase of $\pi$ with particles of charge $1/2$, meaning that they are physically observable, and thus cost a finite energy per unit length.  The charge-$1/2$ excitations (the only possible charges in this limit) are therefore confined.   Indeed, this model is exactly the $p=2$ \AWW model discussed in Sect. \ref{Sec2p}.
 
 Finally, consider $p=8, m=4$.  In this case there are four electric fluxes that remain for $\lambda/\Gamma =0$, of strength $1/4, 1/2,$ and $3/4$.  The corresponding charges are bound to flux tubes of strength $4 \pi, 8 \pi$, and $12 \pi$, respectively.  The first and last of these fluxes have Berry phase $\pi$ with objects of charge $1/4$, and thus are physically observable.  A flux of $8 \pi$ is physically unobservable, so that there is a deconfined excitation of charge $1/2$ (corresponding to $s=4$ in the original spin basis).  According to our criteria above, this excitation is a boson.  The excitations of charge $1/4, 3/4$ are confined.  This model is therefore intermediate between the Potts gauge theories and the \AWW models: it has deconfined charge and non-trivial topological order, but the number of deconfined charges (and the ground state degeneracy) is less than the number of deconfined spin labels.

 \section{Ising gauge transitions in Walker-Wang models for SU(2)$_k$}\label{s:SU(2)}

The \AWW models are only a subset of the possibilities that can be realized following the construction of Ref. \onlinecite{WW}.  In this section and the next, we will discuss transitions in a different family of Walker-Wang models, whose low-lying excitations include non-abelian anyons confined to their surfaces.  
We will call these the SU(2)$_k$ Walker-Wang models, as on lattices with boundaries, their surface states have the topological order of a chiral SU(2)$_k$ Chern-Simons theory.  

In the \suw models, each edge is endowed with a spin variable
$
s = 0, 1/2, ... k/2
$, 
 where $k$ is a parameter of the model.  
The vertex operator acts according to 
\be \label{QVConstr}
 \hat{\mac{Q}}_V\ \  \includegraphics[totalheight=.4in]{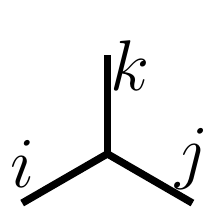} = \sum_{l \in i \times j} \delta_{kl} \ \  \includegraphics[totalheight=.4in]{Vertex.pdf}
\ee
where  
\be \label{Lqadd}
 i \times j \equiv \sum_{l = |i - j|}^{ \text{min} \{ i+j, k-i-j\} } l
 \ee
Notice that if $k = \infty$, we recover the usual rules for the addition of spin angular momenta, and $\hat{\mac{Q}}_V$ has eigenvalue $1$ on states where the total angular momentum is conserved at the vertex $V$.  Eq. (\ref{Lqadd}) enforces a ``deformed" version of this condition, appropriate to models with a finite number of possible values of the total (deformed) spin angular momentum on each edge.\footnote{Needless to say, deforming the rules for adding angular  momentum in this way means that the edge labels can no longer be associated with representations of SU(2); instead they can be associated with representations of the affine algebra SU(2)$_k$.}
 
As in the Abelian models, the plaquette operator $\hat{B}_P$ is a superposition of operators that raise all spins on the edges around the plaquette by $s$ (using the rules (\ref{Lqadd}) ), and simultaneously multiply the wave-function by a configuration-dependent complex coefficient:
\be
\hat{B}_P = \sum_{2s = 0}^{ k} \hat{\Phi}_{P,s} 
\ee
In this case we cannot separate the action into independent raising and phase operators, as in general raising all spins on the edges of $P$ by $s$ will create a superposition of configurations with different spin labels; each element of this superposition may have a different phase.  Thus we denote by $\hat{\Phi}_{P,s}$ the combination of raising operator and phases.  (The precise form of these operators is given in Refs. \onlinecite{WW,WWUs}; see also Sect. \ref{ss:WWgen}).

Following the procedure that we used in the Abelian models above, we will seek to identify possible edge terms (transverse fields) that we can add to the solvable Walker-Wang Hamiltonian to produce a phase transition.  In the Abelian case, the transverse field operators $\oh_e$ were sums of terms $ \oS_{e, +}^n +\oS_{e, -}^n $, each of which has eigenvalues of the form
$ \cos \left( 2 \pi n s_e / p \right )$.  For the SU(2)$_k$ models, it is convenient to choose a slightly different form of the transverse field:
\be \label{hSU2}
\oh_e^{(m)} |s_e \rangle  = \kappa_e \sin \left( \frac{  \pi (2 s_e +1)( 2 m +1 )}{k+2 } \right )|s_e \rangle
\ee
where $s_e, m \in \{ 0, 1/2, 1, ... k/2 \}$, and $\kappa^{-1}_e =\sin \left [ \pi (2 s_e +1)/(k+2) \right ]$.  
This has a similar effect to adding a term of the form $\cos (4 \pi j s /k)$, but assigns slightly different energy penalties to the various edge spins.  The advantage of the form (\ref{hSU2}) is that it obeys the rules (\ref{Lqadd}) for combining spins:
\be \label{LaddS}
\oh_e^{(m)}  \oh_e^{(n)} |s_e \rangle = \sum_{l \in m \times n} \oh^{(l)} |s_e \rangle
\ee
The excitations created by the operator $\oh_e^{(m)}$ is therefore a vortex loop ``of spin $m$", and can be combined according to the same rules (Eq. \ref{Lqadd}) we use to raise and lower the edge spins.

 For most choices of $k, m$, the transverse field  
will assign an energy cost to every non-0 spin, and the condensed phase is the ``trivial phase", in which all spin labels are confined.  (If we choose $m=1/2$, for example, this is true for all $k$).    We will not discuss these transitions in detail, but note that they {\it cannot} be mapped onto the transitions discussed above: the commutator of $\oh_e^{(m)}$ with $\hat{B}_P$ is not a sum of terms of the form (\ref{TheZLikeCommutator}), and the condensing vortex loops do not behave like vortex loops in a $\mathbb{Z}_p^{\text{Potts}}$ (or related Abelian) model. 

There are, however, some choices of the transverse field term which do not lead to a transition into the trivial phase.  These will be our primary interest here.  In Sect. \ref{s:gencat} we will describe a general method to determine the topological order of the condensed phases for general $m$.

 We begin, however, with two examples.  
First, if we take $m =1$,
\be \label{h2Tor}
\oh_e^{(1)} |k/2 \rangle
  = \kappa_1  |k/2 \rangle \ \ \ \ \ \ \ \ \  \oh_e^{(1)} |0 \rangle
  = \kappa_1  |0 \rangle
\ee
with $\kappa_1 = \sin \left( \frac{ 3 \pi}{ k+2 } \right)$.  It is not hard to show that the eigenvalue of $\oh_e^{(1)} $ for all other spin states is smaller, so that this transverse field assigns an energy penalty to any edge label {\it except} $0, k/2$.
In this case, we will have little to say about the behaviour near the transition.  However, we will show that the condensed phase is (1) the $\mathbb{Z}_2$ gauge theory if $k$ is divisible by $4$; (2) the $\mathbb{Z}_2$ gauge theory with fermionic charges if $k = 4 n +2$; (3) the $p=2$ \AWW model (Sect. \ref{Sec2p}) if $k$ is odd.  The difference between these cases arises from the different possible phases incurred by acting with $\hat{\Phi}_{P, k/2}$ on configurations with $s_e =0, k/2$ only, as we discuss in Sect. \ref{s:gencat}. 

A second interesting possiblity is to take $m = k/2$.  In this case we have
\be \label{k2Eq}
\oh_e^{(k/2)} |s_e \rangle  = - \cos \left( (2 s_e +1) \pi \right )  |s_e \rangle = (-1)^{2 s_e} |s_e \rangle
\ee
such that $- \oh_e^{(k/2)} $  assigns an energy penalty to any state where the edge $e$ carries a half-integer spin.  This case is the 3D analogue of the Ising transitions studied by Ref. \onlinecite{TSBLong, TSBShort}.  We will show that as in the 2D case, for every $k$ the phase transition is identical that of the $\mathbb{Z}_2$ gauge theory.  The nature of the condensed phase, however, depends strongly on $k$, as we shall see.  

 \subsection{Transitions in SU(2)$_2$}\label{ss:SU(2)_2} 
 We begin by studying the simplest model in this family, the SU(2)$_2$ Walker-Wang model.  Here there are only three allowed spin labels $s = 0, 1/2,1$, and since $k/2 =1$, the two choices (\ref{h2Tor}) and (\ref{k2Eq}) of transverse field coincide.  
 
 The Hamiltonian we will study is
 \ba
 H = \frac{1}{2} \Lambda \sum_P (\mathbf{1} -  \hat{\Phi}_{P,1} ) - M \sum_V \oQ_V  \n
 + \frac{ \lambda}{\sqrt{2}} \sum_P  \hat{\Phi}_{P,1/2} - \Gamma \sum_e \oh_e^{(1)}
 \ea
with the action of $\oh_e$ given by Eq. (\ref{k2Eq}).  

The action of the vertex term is:
\be
\oQ_V | \includegraphics[height=.35in]{Vertex.pdf} \rangle = \begin{cases} 1 \ \ \ &
\includegraphics[height=.35in]{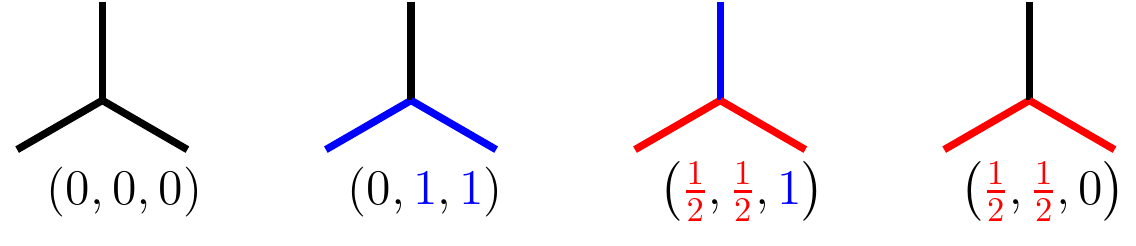} \\
0 & \text{otherwise} 
\end{cases}
\ee
where it is understood that rotations of these combinations have the same eigenvalue under $\oQ_V$.
Since the eigenvalue of $\oQ_V$ is conserved at every vertex, we will restrict our attention to states where these eigenvalues are $1$ everywhere.  In this restricted Hilbert space, edges of spin-$1/2$ always occur in closed loops, while edges of spin-$1$ can either form closed loops, or open lines ending at vertices with two spin-$1/2$ edges.  

We have separated the action of the plaquette operator into two terms:
\be
\lambda( 1 -  \hat{B}_P )  \rightarrow \frac{1}{2} \Lambda  \left( \mathbf{1} - \hat{\Phi}_{P,1}  \right) +\frac{\lambda}{ \sqrt{2}} \hat{\Phi}_{P, 1/2} 
 \ee
 The term proportional to $\Lambda$ changes the spin on each edge of the plaquette $P$ by an integer amount, and thus commutes with the transverse field term (and in fact with all terms in the Hamiltonian).  The term proportional to $\lambda$ changes the spin by a half-integer amount -- and therefore changes the eigenvalue of $\oh_e$ on all edges of $P$.  (Walker and Wang's\cite{WW} definition of the plaquette operator $\hat{B}_P$ is obtained by setting $\Lambda = \lambda$.)
 
In more detail, the action of the operators $\hat{\Phi}_{P, s}$ is as follows. 
$
\hat{\Phi}_{P,1}$ raises the spin on every edge of $P$ by $1$, using the rules (see (\ref{Lqadd}) ) 
\be
1 \times 0 = 1 \ \ \ \ \ \  1 \times 1 = 0 \ \ \ \ \ \ \ \ 1 \times 1/2 = 1/2
\ee
In other words, spin$-1/2$ edges remain spin-$1/2$ edges, while edges of spin-$0$ and spin-$1$ are interchanged.  (The operator $\hat{\Phi}_{P, 1}$ also in general multiplies the wave-function by a non-zero configuration-dependent coefficient; we will have more to say about these coefficients presently).  
$\hat{\Phi}_{P,1/2}$ raises the spin on every edge of $P$ by $1/2$, using the rules (see (\ref{Lqadd}) ) 
\be
1/2 \times 0 = 1/2  \ \ \ \ \ \  1/2 \times 1 = 1/2 \ \ \ \ \ \ \ \ 1/2 \times 1/2 = 0 + 1
\ee
In other words, it turns edges of spin $0$ or $1$ into edges of spin $1/2$, and edges of spin $1/2$ into a superposition of edges of spin $0$ and spin $1$.  Again, this action also multiplies the result by a configuration dependent coefficient.  Importantly, this coefficient is $0$ for any configurations where the eigenvalue of $\oQ_V$ has changed.  This ensures that the plaquette and vertex terms commute.

 We now wish to prove the two assertions following Eqs. (\ref{h2Tor}) and (\ref{k2Eq}): first, that this transition is identical to the confining transition of the $\mathbb{Z}_2$ gauge theory, and second, that the condensed phase at large $\Gamma/\lambda$ can be described by the {\it deconfined} phase of $\mathbb{Z}_2$ gauge theory with fermionic sources.  
 
 We begin with the second of these assertions.  As in previous sections, to understand the behavior of the condensed phase,  it suffices to consider the Hamiltonian at $\lambda=0, \Gamma >0$.  In this case the spin on each edge is conserved modulo an integer, and each edge of spin-$1/2$ carries an energy cost of $\Gamma$: spin-$1/2$ labels are confined, in the sense described in Sect. \ref{PottsSec}.  
Restricting our attention to states containing only the deconfined spin-$1$ label, the effective Hamiltonian has the form
\be \label{HSpecial}
 H = \frac{1}{2} \Lambda \sum_P (\mathbf{1} -  \hat{\Phi}_{P, 1}  ) - M \sum_V \oQ_V 
 \ee
where $\oQ_V$ penalizes states with an odd number of spin-$1$ edges entering the vertex $V$.  In the restricted Hilbert space, $\hat{\Phi}_{P,1}$ simply interchanges all spin $0$ edges of $P$ with spin $1$ edges, and all spin $1$ edges with spin $0$ edges, multipliying the result by a coefficient $\pm 1$.  Apart from a possible sign in the action of the plaquette term, we recognize this as the Hamiltonian for a \gz gauge theory, whose possible excitations are vortex loops of plaquettes on which $ \hat{\Phi}_{P, 1} =0$, and vertex violations (charges) where an odd number of $1$ spins meet at a vertex.  As we mentioned in Sect. \ref{GenAWWSect}, the fact that the matrix elements of the plaquette term are always real ensures that the charges are deconfined; the effect of the $-$ signs in this case is to render the charges fermionic, as we prove
 in Appendix \ref{FermiApp}.  
 
 The Hamiltonian (\ref{HSpecial}) is, of course, valid only precisely at $\lambda=0$; however, the qualitative features of the \gz   gauge theory describe the low-energy physics of the entire phase.  Since spin-$1/2$ edges are confined throughout the large $\Gamma/\lambda$ phase, these loops can never appear at long length scales, and Eq. (\ref{HSpecial}) is the effective Hamiltonian at long-wavelengths throughout this phase.
 
Having understood the condensed phase, let us understand the nature of the transition.  This is most easily done if we restrict our attention to states for which 
\be \label{EqRestricted}
\oQ |\psi \rangle = |\psi \rangle \ \ \ \ \ \ \ \ \  \frac{1}{2}\left( \mathbf{1} + \hat{\Phi}_{P, 1}  \right) |\psi \rangle = |\psi \rangle
\ee
i.e. states with no vertex violations, and no spin-$1/2$ vortices.  ($\hat{\Phi}_{P, 1}$ does not measure the spin-$1$ vortices generated by $\oh_e^{(1)}$).
In this subspace, for $\Gamma =0$ the model contains only one type of excitation: a plaquette defect visible only to $\hat{\Phi}_{P, 1/2}$, which costs energy $\lambda$.  This is precisely the plaquette defect created by the transverse field term $\oh_e^{(1)}$, which commutes with $\hat{\Phi}_{P, 1}$ and anti-commutes with $\hat{\Phi}_{P, 1/2}$.  Further,  $\left( \oh_e^{(1)}\right)^2= \mathbf{1}$, exactly as for the transverse field term in the $\mathbb{Z}_2$ gauge theory.  Thus we might expect that the states in the Hilbert space defined by Eq. (\ref{EqRestricted}) can be mapped onto states of the pure $\mathbb{Z}_2$ gauge theory, by identifying the vortex loops created by $\oh_e^{(1)}$ with the vortex loops in the gauge theory.  

It is instructive to construct such a mapping directly in the spin eigenbasis.  On each edge, we may map the degrees of freedom of the SU(2)$_2$ model to that of a \gz gauge theory via:
\be \label{s2s}
s_e \rightarrow 2 s_e \ \ \ \ (\text{ mod } 2)
\ee
The reason that this is a sensible mapping is that at each vertex $V$, the number of half-integer spins entering $V$ must be even; similarly in the \gz gauge theory the number of edges with $s=1$ must be even.  Thus in the absence of vertex violations, the mapping is one from closed loops of spin $1/2$ to closed loops of spin $1$.  Further, the transverse field term $\oh^{(1)}_e = (-1)^{2 s_e}$ maps exactly to the transverse field term $\oh_e = (-1)^{s_e}$ of the \gz gauge theory.  Finally, the operator $\hat{\Phi}_{P,1/2} $ interchanges integer and half-integer spins, while the plaquette term in the \gz gauge theory interchanges $s_e =0$ and $1$.  

The mapping (\ref{s2s}), however, clearly fails to capture many of the features of the ground state of the SU(2)$_2$ Walker-Wang model.   First, as a mapping of spin configurations it is many-to-one: for a given choice of spin-$1/2$ loops, there are many possible ways to occupy the remaining edges of the lattice with spin $0$ or $1$ states without violating the vertex condition.  Second, the matrix elements of $\hat{\Phi}_{P, 1/2}$ are not all equal in magnitude, as is the case for the \gz gauge theory.  Third, matrix elements of the plaquette term in the \gz gauge theory are also real and positive, while for SU(2)$_2$ they can be negative or complex.

It is thus somewhat surprising that the expectation values of all operators required to describe the phase transition are the same in both models.  Essentially, this occurs because operators that commute with the conditions (\ref{EqRestricted}) can do only one of two things: either they measure the spin on each edge mod $1$ (and are thus a linear combination of transverse field operators), or they are linear combinations of $\hat{\Phi}_{P,1}$ (which in any case must act as the identity on our restricted Hilbert space) and $\hat{\Phi}_{P,1/2}$.  
As we show in Appendix \ref{PhaseTranApp}, the fact that the commutators of these operators have the same structure as their analogues in \gz gauge theory ensures that their expectation values are identical in both models.  

For example, for $\Gamma=0$ any such expectation value can be expressed in terms the expectation value of an operator that is diagonal in the the spin basis, after moving all  $\hat{\Phi}_{P,1/2}$, $\hat{\Phi}_{P,1}$ terms to the right of all $\oh_e$ terms using the commutation relations, and using the fact that $\hat{\Phi}_{P,1/2} |\Psi_0 \rangle_{\Gamma=0} = \hat{\Phi}_{P,1} |\Psi_0 \rangle_{\Gamma=0}=|\Psi_0 \rangle_{\Gamma=0}$.  Expectation values of operators diagonal in the spin basis are insensitive to the relative phases between different spin configurations in the ground states of the two models; they  depend only on the probability of being in a given loop configuration.  As we discuss in Appendix \ref{PhaseTranApp},  in the SU(2)$_2$ model adding  (or removing) a closed loop of spin-1/2 edges  halves ( or doubles) the number of possible integer spin configurations compatible with the given choice of spin-1/2 loops.  However, this is exactly compensated for by the fact that the coefficients in the action of the plaquette projector multiply such configurations by a factor of $\sqrt{2}$ (or $1/\sqrt{2})$ relative to the state from which they were derived.  Hence the {\it probability} to be in a given loop configuration, and hence the expectation value of any product of transverse field terms in the $\Gamma=0$ ground state, is the same in both models.

\subsubsection{\gz Transitions in SU(2)$_k$}

$k=2$ is a special case, in which the condensed phase is a Potts gauge theory, {\it and} the transition is also of the form described in Sect. \ref{Sec2p}.  For larger values of $k$ this is no longer the case.  Here we briefly comment on the more general situation.

First, we have already observed that we can always add a transverse field operator of the form $\oh_e^{(k/2)} = (-1)^{2 s_e}$.  For transitions where this is the only transverse field, if we restrict our attention to states satisfying
\be \label{ER2}
\oQ |\psi \rangle = |\psi \rangle \ \ \ \ \ \ \ \ \  \frac{1}{\mac{D}}\left( \mathbf{1} + \sum_{s = 1}^{\lfloor k/2 \rfloor} \hat{\Phi}_{P, s} \right) |\psi \rangle = |\psi \rangle
\ee
(where $\mac{D}$ is an appropriate normalization, such that the operator has eigenvalues $0$ and $1$)
the mapping (\ref{s2s}) between the low energy degrees of freedom in the SU(2)$_k$ model and the \gz theory remains valid.  The rules for allowed combinations of angular momenta at a vertex ensure that edges of half-integer spin form closed loops, allowing us to map from closed 1/2-integer spin loops to closed integer spin loops in the \gz gauge theory.  Here again, because the only excitations relevant to the transition are vortex loops created by $\oh_e^{(k/2)}$ -- which are Ising-like, as the operator squares to the identity -- one can show that all correlation functions in the limit (\ref{ER2}) are equal to their \gz analogues.  This is discussed in more detail in Appendix \ref{PhaseTranApp}.

The solvable Hamiltonian describing the physics of the condensed phase is simply 
\be
H = M \sum_V \oQ_V+ \Lambda \sum_P \left[ 1 -  \frac{1}{\mac{D}}\left( \mathbf{1} + \sum_{s = 1}^{\lfloor k/2 \rfloor} \hat{\Phi}_{P, s} \right) \right]  
\ee
The Hilbert space in the solvable limit contains only integer spins on the edges.  For $k>2$, however, the vertex condition on these integer spins is not that of a \zp gauge theory: for example, if $k=4$, we have
\be
1 \times 1 = 0 + 1 + 2
\ee
and vertices with 3 spin $1$ edges, or two spin $1$ edges and a spin $2$ edge, are allowed.  We will see in the next section how to determine the topological characteristics of these condensed phases.

 \section{Topological order and confinement in non-Abelian Walker-Wang models}\label{s:gencat}
Having investigated the confining phase transitions in Abelian and SU$(2)_2$ Walker-Wang models, we now show that many of the results found follow naturally from the mathematical structures (`pre-modular categories') used to define Walker-Wang models. While the methods of this section will not furnish us with the details of the phase transition, they do tell us how topological order changes for {\it any} confining transition in a Walker-Wang model. Before setting out a general recipe, we summarise the basic properties of general WW models and their corresponding categories in \secref{ss:WWgen} (see [\onlinecite{WWUs}] for details). Then in \secref{ss:Wmatrixtransitions} we show that the results of \secref{GenAWWSect} and \secref{ss:SU(2)_2} are simply described in terms of the $M$-matrix (closely related to the $S$-matrix) of the categories corresponding to the \AWW and SU$(2)_k$ Walker-Wang models respectively. Following this, we describe the condensation in SU(2)$_k$ for the cases $k>2$ which did not yield to the methods of \secref{s:SU(2)}.

\subsection{General Walker-Wang models and the $M$-matrix}\label{ss:WWgen}
\begin{figure}
\begin{center}
\includegraphics[width=.975\linewidth]{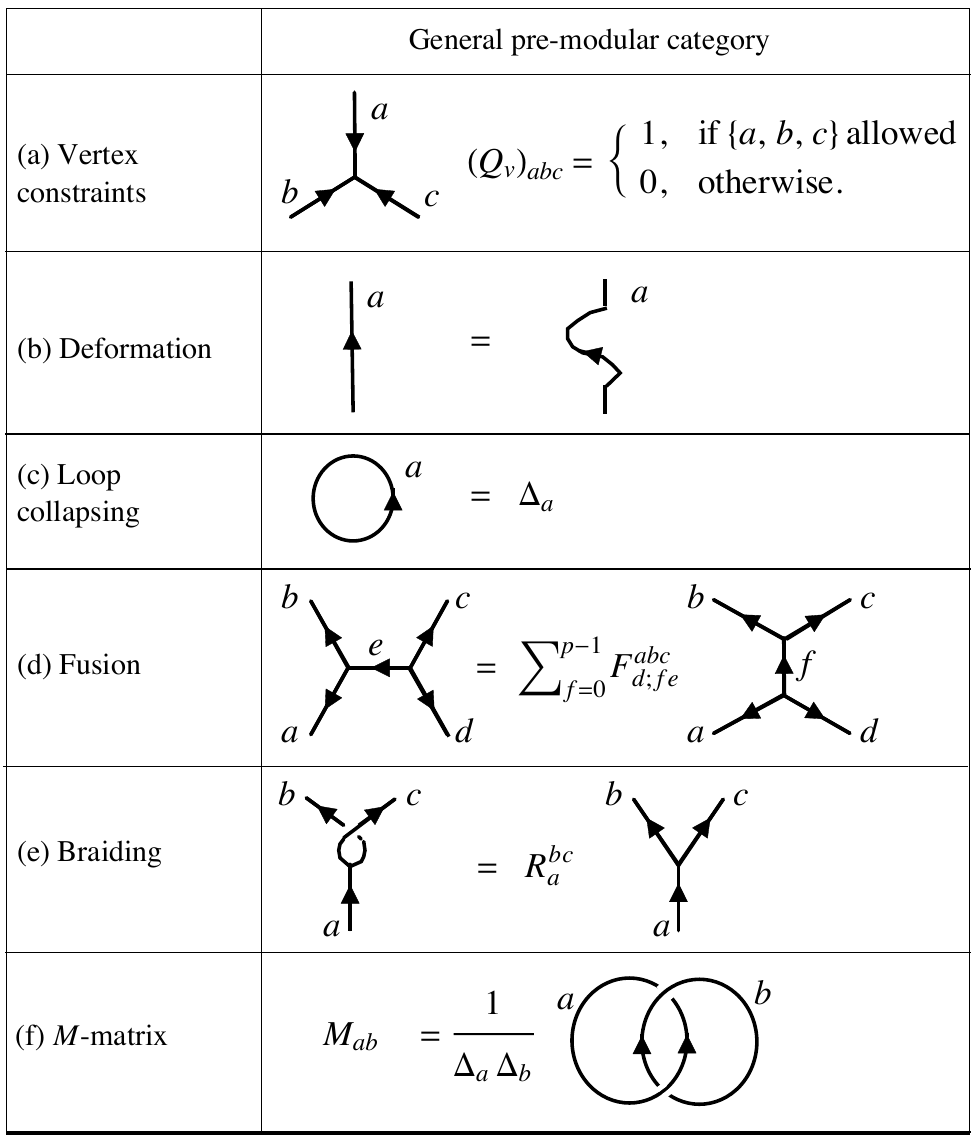}\\
                \caption{(a) Represents the vertices allowed by the category; the ground state of a Walker-Wang model will involve only these types of vertices. The diagrams in (b)-(d) serve two purposes. Firstly, they tell us the relative amplitudes of string-net configurations in the ground state e.g. row (c) tells us that configurations related by removing a closed loop carrying label $a$ occur with a relative factor of $\gD_a$ in the ground state. Second, these diagrams provide a neat graphical mnemonic for the definitions of string operators. Note that there is a rule conjugate to (e) obtained by turning the over-crossing into an under-crossing on the left hand side, and sending $R \ra R^{*}$ on the right hand side. }
                \label{fig:gencat}
\end{center}                
\end{figure}

General WW models have a Hilbert space consisting of a $p$ state system on each edge of the lattice shown in Fig. \ref{LatticeFig}. We denote the $p$ possible states by the labels $\{a_0 ,a_1,\ldots,a_{p-1}\}$, where $a_0$ is the identity element; for the models of \ref{AWWSec} these labels were the possible spins $s_e =0, 1, ... p-1$ with $s_e=0$ corresponding to $a_0$.   Each label in the set has a conjugate label also in the set, which we denote $\overline{a}_i$.  (For the abelian models, $\overline{s} = p-s$, while for SU(2)$_k$, $\overline{a} = a$).   As we saw in \secref{AWWSec}, if $a \neq \overline a$, to define $\hat{B}_P$ and $\oQ_V$ we must include arrows on each edge to specify whether the operator measures $a$ or $\overline{a}$. The general Hamiltonian has the familiar form
\be\label{eq:HgenWW}
H = \gl \sum_P (1-\hat{B}_P) - M \sum_{V}\oQ_V - \Gamma \sum_{e} \oh^{(m)}_e
\ee
where $\oh^{(m)}_e$ is defined below, and the operators obey the commutation relations in equation \eqref{WWCom1} and \eqref{WWCom2}. For now we tune the model to an exactly solvable point by setting $\Gamma=0$. The $\oQ_V$ term ensures the ground state is a superposition of configurations for which only certain combinations of the $p$ labels are allowed to meet at each vertex.  The allowed combinations are those that satisfy the {\it fusion rules} of the category, as in \figref{fig:gencat} (a) -- we call configurations that satisfy this condition at every vertex `string-net' configurations. 

The plaquette term $\hat{B}_P$ is subtler. Just as in the abelian case detailed in \eqnref{BpPotts}, plaquette violations can be labeled by their `fluxes', which are drawn from the same set  $\{a_0,a_1,\ldots,a_{p-1}\}$ as the edges (see \secref{ss:AWW:Ham}).   These fluxes can be measured by a plaquette operator analogous to $\hat{\Phi}_{P}$, which raises or lowers the edge labels on the boundary of the plaquette $P$ according to the rules of the category (see Refs.~\onlinecite{WW,WWUs} for more detail). Since $\hat{B}_P$ and $\oQ_V$ commute, the result is that the ground state is a superposition of different string-net configurations, with coefficients that are related by the set of graphical rules in \figref{fig:gencat}(b)-(e). Only certain combinations of graphical rules in \figref{fig:gencat}(a)-(e) are consistent with one another; when they are consistent, the resulting structure is called a premodular category. 

The $p\times p$-sized `Monodromy' or $M$-matrix (which is essentially the modular $S$-matrix with a different normalization) shown in \figref{fig:gencat} contains much of the useful information in the category. We will use this object to ascertain the topological order of phases on either side of the confining transition. Before doing so, we note that it satisfies\cite{BondersonThesis}
\be\label{Mmatrixproperties}
\abs{M_{ab}} \leq 1 \ \ \ \ M_{a 0} = 1 \ \ \forall  a  \in \{a_0,a_1,\ldots,a_{p-1}\} \punc{.}
\ee
In Ref.~\onlinecite{WWUs}, we showed that  the spectrum of point defects (and the associated topological order) can be deduced from the $M$-matrix (or the closely related $S$-matrix) of the category.  Our main result was that point defects carrying charge $j\neq 0$ are deconfined in the bulk if and only if $M_{ ij}  = 1$ for all possible labels $i$\footnote{In Ref.~\onlinecite{WWUs} we showed that a WW model has deconfined bulk excitations if and only if the $S$-matrix is degenerate. Precisely the same statement holds for the $M$-matrix, because $M$ is obtained from $S$ by rescaling rows and columns.}. As an example, we now use the $M$-matrix to shed light on Walker-Wang models in the $p=2$ models  discussed in \secref{Sec2p}. The $p=2$ Potts gauge theory is a Walker-Wang model based on the so-called $\integ^{(0)}_2$ category, while the $p=2$ \AWW model is based on a category $\integ^{(1/2)}_2$. Both categories have edge labels in $\{0,1\}$, and fusion rules specifying that an even number of label-$1$ edges must enter each vertex. The full properties of these categories is listed elsewhere\cite{BondersonThesis}, but here we require just their $M$-matrices
\be
M_{\text{T.C.}} =  \left(\begin{array}{cc}1 & 1 \\1 & 1\end{array}\right) \ \ \ M_{\text{3DSem}} =\left(\begin{array}{cc}1 & 1 \\1 & -1\end{array}\right)\punc{.}
\ee
In the case of the $p=2$ Potts gauge theory (or ``3D toric code''\cite{HammaPRB72,CastelnovoPRB78}), we see that both columns of $M$ are filled with $1$, so that particles carrying label $1$ are deconfined. On the other hand the $M$-matrix of the $p=2$ \AWW model (also known as the ``3DSem model''\cite{WWUs}) has only its first column filled with $1$, and so there are no non-trivial deconfined particles in the bulk of this model. In the less trivial non-abelian example studied in \secref{ss:SU(2)_2}, the Walker-Wang model based on SU(2)$_2$ has labels in $\{0,\frac{1}{2},1\}$ and an $M$-matrix
\be
M_{\text{SU(2)$_2$}} = \left(\begin{array}{ccc}1 & 1 & 1 \\ 1 & 0 &-1 \\1 & -1 & 1\end{array}\right)\punc{.}
\ee
Thus the $SU(2)_2$ model from \secref{ss:SU(2)_2} has no deconfined bulk species since no column except the first is all unity. Having stated the correspondence between the $M$-matrix and the spectral properties of the corresponding Walker-Wang models, we now use it to understand the condensation transitions between different models. 

	\subsection{Confining transitions from the $M$-matrix}\label{ss:Wmatrixtransitions}
Ramping up $\Gamma$ in \eqnref{eq:HgenWW} drives the system through a confining transition, where loops proliferate. In this section, we will describe the effect of this transition on the spectrum of the theory. To begin, pick an edge $e$, and suppose it is labelled $j$ in the $\hat{n}$ direction. To create a small vortex loop of flux $m$ encircling $e$ in a right-hand sense with respect to $\hat{n}$, define
\be
\hat{v}^{(m)}_e\ket{ j } = M_{m j} \ket{ j }\punc{.}
\ee
In the simplest case, starting with the $\Gamma=0$ ground state, the state formed by $\hat{v}^{(m)}_e\ket{\text{GS}}$ has $0$ flux on every plaquette except those bordering $e$. If we set $\oh^{(m)}_e=(\hat{v}^{(m)}_e+(\hat{v}^{(m)}_e)^{\dagger})/2$, then the resulting operator creates a superposition of flux $m$ and flux $-m$ loop around the edge $e$. Using the properties of the $M$-matrix in \eqnref{Mmatrixproperties} it is clear that the theory in the $\Gamma \ra \infty$ limit consists of a string-net with edge labels $j$ satisfying $M_{mj}=1$, although we are unable to say precisely how the system behaves for intermediate $\Gamma$. 

Starting with a category with labels $\{a_0 ,a_1,\ldots,a_{p-1}\}$  and proliferating vortex loops of flux $m$, we can use the $M$-matrix to answer most questions about the topological order of the condensed phase. Firstly, the labels present in the condensed phase are precisely $\{\wa_{0},\wa_{1}, \ldots, \wa_{l-1} \}$ such that $M_{\wa_{i} m} = 1$. The model obtained in the $\Gamma\ra\infty$ limit is a new Walker-Wang model based on a category which has has the same graphical rules as the old category, except the rules are now restricted to labels $\{\wa_{0},\wa_{1}, \ldots, \wa_{l-1} \}$ i.e., this is now a subcategory of the original category. In particular the new $M$-matrix is simply the $l\times l$ matrix $\widetilde{M}_{i j} = M_{\wa_i \wa_j}$. Rather non-intuitively, if particles carrying label $\wa$ are confined for small $\Gamma$ (i.e. $M_{\wa a_i}\neq 1$ for some $a_i$) they can be deconfined in the condensed (large $\Gamma$) phase, since it is possible that $M_{\wa \wa_i}= 1$ for the $l$ labels $\wa_i$ present in this phase. 

\subsubsection{Abelian categories}
We now reproduce some of the results of \secref{GenAWWSect} using the $M$-matrix formalism. The \AWW models of interest have a $p$-state system on the edge of each lattice, and the transition in \secref{GenAWWSect} was  driven by proliferating vortex loops with an equal superposition of flux labels in $\{m,2m,\ldots, p-m\}$ (see \eqnref{hm}). However, proliferating only the $\pm m$ vortex loops gives the same ground state in the $\Gamma\ra \infty$ limit (though the behavior at intermediate $\Gamma/\lambda$ may differ).  To parallel more closely our treatment of the non-abelian models we will use this alternative form for the transverse field here, as it  reproduces the topological orders found in \secref{GenAWWSect} for the condensed phase. 

In the uncondensed phase,  the $M$-matrix takes the form $M_{a b} = e^{\pm i 2 \pi a b/ p}$ with $a,b \in \{0,1,\ldots,p-1\}$, where we take  $\pm$ for $p$ even or odd respectively. If $m$ labeled vortices are condensed, edges with label $a$ develop a gap of $\sim \Gamma(1-\text{Re}\ds{M_{a m}})\geq 0$, so the labels surviving on the edges of the lattice are precisely those satisfying $M_{a m} = 1$ i.e. $\{0, p/m, 2p/m, \ldots, (m-1)p/m \}$. It follows readily that as $\Gamma\ra \infty$, the ground state of \eqnref{eq:HgenWW} is precisely that of \secref{s:OtherAbelianTrans} in the same limit. Moreover, this ground state can be thought of as arising from a Walker-Wang model with the same graphical rules as in the uncondensed phase, but restricted to the surviving labels. The effective $M$-matrix of the new model is an $m \times m$ matrix of the form  $\widetilde{M}_{x y} = e^{\pm i 2 \pi x y p / m^2}$, with rescaled labels $x,y\in \{0,1,\ldots,m-1\}$, and we can use this to diagnose the topological order of the vortex phase. We now summarise the cases presented in \tabref{tab:CAWW} using this new formalism: 

\begin{itemize}
\item$p/m^2=2n$: Clearly $\widetilde{M}_{x y}=e^{\pm i 4 \pi x y n}=1$ for all $x,y$, so that all $m$ remaining edge labels are deconfined. Therefore the model has the topological order of a $\integ_m$ gauge theory. A pair of particles carrying label $x$ have mutual bosonic statistics because the $R$-matrix (\figref{fig:gencat}(e)) of the category satisfies $R_{2x}^{x x} =e^{\mp i 2 \pi x^2 n}=1$ (see \appref{FermiApp}). 

\item $p/m^2=2n+1$: Again $\widetilde{M}_{x y}=e^{\pm i 2 \pi x y (2n+1)}=1$, so the model appears to have the topological order of a $\integ_m$ gauge theory. However in this case a pair of particles with odd charges $x$ exhibit fermionic statistics, a fact that follows from the form of the $R$-matrix $R_{2x}^{x x} =e^{\mp i  \pi x^2 n}=(-1)^x$ (see \appref{FermiApp}).

\item$p/m^2=n/m$ an irreducible fraction: In this case $\gcd(p/m,m)=1$, which implies that $\widetilde{M}_{x y} = e^{\pm i 2 \pi x y n/ m}$ is only ever equal to $1$ when $xy=0 \mod m$, which in turn implies that all  point particles are confined in the bulk. Thus the resulting model is a \AWW, and only has surface (rather than bulk) topological order.

\item$p/m^2=n/b$ with $1<b<m$: Letting $n/b$ be a fraction in lowest terms implies that $\widetilde{M}_{x y} = e^{\pm  i 2 \pi x y n/b}$. Clearly $m/b$ columns of $\wM$ are filled with $1$, which implies there are $(m/b)-1$ (non-trivial) deconfined species, and $m(1-1/b)$ confined species. Thus the condensed phase has both bulk topological order, and surface topological order. 
\end{itemize}
\subsubsection{SU$(2)_k$}
 
Having seen that the results of \tabref{tab:CAWW} are reproduced precisely by the $M$-matrix formalism, we now investigate a non-abelian example. The category $SU(2)_k$ has $k+1$ labels $\{0,\frac{1}{2},1,\ldots, \frac{k}{2}\}$, and the $M$-matrix takes the form\cite{BondersonThesis}
\be\label{eq:SU(2)_k}
M_{a b} =\frac{ \sin \db{\frac{\db{2a+1}\db{2b+1}\pi}{k+2}} \sin\db{\frac{\pi}{k+2}} }{ \sin \db{\frac{\db{2a+1}\pi}{k+2}} \sin \db{\frac{\db{2b+1}\pi}{k+2}}}\punc{.}
\ee
We now summarise the results of proliferating label $m$ vortices.
\paragraph*{Condense half-integer $m$:}
For half-integer $m$, one can check that $M_{mj}<1$ unless $j=0$. Therefore, proliferating half-integer vortices $m$ confines all nontrivial edge labels, leading to the trivial string-net with only the zero label in the $\Gamma \ra \infty$ limit. 
\paragraph*{Condense integer $m$ with $0<m < \frac{k}{2}$:} In this case, $M_{m j}<1$ unless $j= 0,\frac{k}{2}$. Hence proliferating $m$ label vortices gives a string-net phase with two labels $ 0,\frac{k}{2}$. The effective $M$-matrix for these remaining particles takes the form:
\be
\wM \propto \left(\begin{array}{cc}1 &1\\
1				     & (-1)^{k}\end{array}\right)
\ee
Therefore, in the case that $k$ is even, the resulting phase has the topological order of a $\integ_{2}$ gauge theory (or equivalently the 3D toric code). In the case that $k$ is odd, the phase has the `surface topological order' of the 3D semion model i.e. the $p=2$ \AWW model.

\paragraph*{Condense $m= \frac{k}{2}$, with $k$ odd:} In the case that $k$ is odd, proliferating the $\frac{k}{2}$ vortex loops leads to a string-net state with only the integer labels $\{0,1,\ldots,\frac{k-1}{2}\}$. This is because $M_{j \frac{k}{2}}<1$ unless $j$ is an integer. Specifying integer particle labels $a=x,b=y$ leads to an effective $M$-matrix:
\be
\widetilde{M}_{x y } = \frac{\sin \db{\frac{\db{2x+1}\db{2y+1}\pi}{k+2}} \sin \db{\frac{\pi}{k+2}} }{\sin \db{\frac{\db{2x+1}\pi}{k+2}}\sin \db{\frac{\db{2y+1}\pi}{k+2}}}\punc{,}
\ee
where $x,y \in \{0,1,\ldots,w-1\}$. This new $M$-matrix has $\widetilde{M}_{x y}<1$ for all $x,y>0$, and therefore all excitations are confined in the vortex phase. The condensed phase is described by a category called SO$(3)_k$\cite{BondersonThesis}.

\paragraph*{Condense integer $m =  \frac{k}{2}$:}In the case the $k$ is even, we again find that $M_{j \frac{k}{2}}=1$ precisely when $j\in\{0,1,\ldots,\frac{k}{2}\}$, hence proliferating vortices with flux $\frac{k}{2}$ leads to a string-net phase with precisely these integer labels.  In this case, however, two columns of the new $M$-matrix (obtained by restricting \eqnref{eq:SU(2)_k} to integer labels) are formed entirely of $1$'s.  These columns correspond to the $0,\frac{k}{2}$-labeled particles, which are the only deconfined particles. We saw an example of this transition in the $k=2$ case where we proliferated the label $1$ vortex, and we were left with $\mathbb{Z}_2^{\text{Potts}}$ topological order.

\subsubsection{SU$(N)_k$}

Among the possible condensation transitions in SU(2)$_k$, therefore, are two notable families: first, by condensing an ``integer spin" vortex loop, we arrive at a condensed phase described by a $\mathbb{Z}_2$ model,
which has bulk or surface topological order if $k$ is even or odd respectively.  Second, condensing the ``highest spin" ($k/2$) vortex loop generically results in a non-abelian topological phase which also has bulk (surface) topological order if $k$ is even (odd). Moreover, this transition has the same low-energy description as the $\mathbb{Z}_2^{\text{Potts}}$ transition in \secref{PottsSec}.

Interestingly, an analogue of both of these exists in SU(N)$_k$ models: proliferating vortex loops in the adjoint representation leaves a set of deconfined edge labels with $\mathbb{Z}_N$ fusion rules. Further, there is always an order $N$ simple current; condensing the corresponding species of vortex loops produces a transition identical to that of the $\mathbb{Z}_N^{\text{Potts}}$ gauge theory.
Though we will not derive these results here (see \onlinecite{DiFrancesco} for the necessary information about these categories), intuitively both families result from the fact that the group SU$(N)$ has a centre $\integ_N$, which is also present in the related tensor category.   The two families of transitions correspond either to condensing the vortices associated with this $\mathbb{Z}_N$ subgroup, or condensing vortices that have trivial Berry phase only with the $\mathbb{Z}_N$ subgroup.

\section{Conclusion} \label{Conclusion}
In this work we have compared a family of phase transitions in the relatively well-studied \zp Potts gauge theories with a related family in the recently introduced Walker-Wang models. In both models, the transitions that we consider can be understood as the condensation of vortex loops (i.e. loops of plaquette defects). Both admit an identical mathematical description of the transition -- allowing us to deduce from the work of Ref. \onlinecite{PottsGauge1} that all of the corresponding phase transitions are first order.   For the Abelian models, ``clock"-like variants of both Walker-Wang and lattice gauge models exist; in these models for $p \geq 5$ the single first-order transition splits into two second-order transitions separated by a gapless Coulomb phase\cite{ElitzurPRD19,HornPRD19,YukawaGuthPRD21}.  

\begin{figure}
 \includegraphics[width=0.975\linewidth]{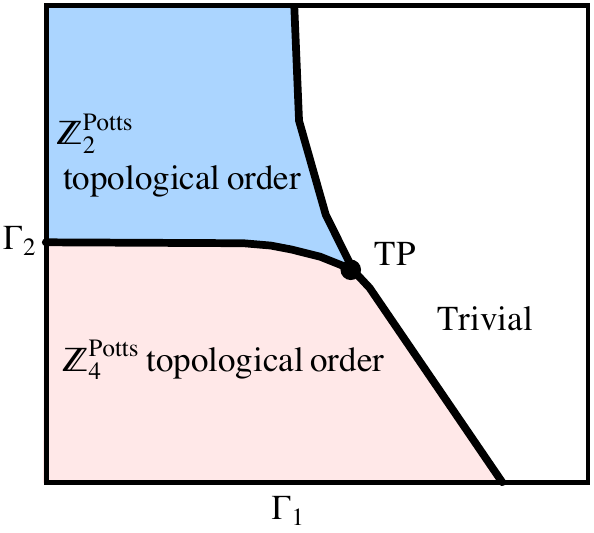}
 \caption{ (Color online) A sketch of the phase diagram for the $\integ_{4}^\text{Potts}$ model, as a function of the transverse field strengths $\Gamma_1$, ($2 \pi$ vortex loops) and $\Gamma_2$ ($4 \pi$ vortex loops), where charge is quantized in multiples of $1/4$ (mod 1).  For small $\Gamma_1, \Gamma_2$, there is a phase with $\integ_{4}^\text{Potts}$ topological order.  This is separated by first-order phase transitions from both the trivial phase at large $\Gamma_1$ and a phase with $\integ_{2}^\text{Potts}$ topological order at large $\Gamma_2$.  For large $\Gamma_1$ and $\Gamma_2$, there is an additional first-order phase boundary between these two condensed phases.  The three first-order lines meet at a triple point.   The location of the phase boundaries is a rough guide, based on numerical analysis of a very similar model by Ref. \onlinecite{CreutzNPB83}.  
 }
\label{Z4Fig}
 \end{figure}

However,
 the relationship between the topological orders of phases connected by such condensation transitions is fundamentally different in the gauge theories and the Walker-Wang systems.
The uncondensed (or deconfined) phase of the $\integ_{p}^\text{Potts}$ gauge theory is a topologically ordered phase with $p$ deconfined charges (i.e. vertex excitations) -- one for each possible value of the electric flux (i.e. edge spin label).  Depending on the value of $p$, there may be several possible condensation transitions.  For any $p$, we can simultaneously condense all magnetic fluxes.  This confines all electric flux lines, thereby confining all charges and completely destroying the topological order.  If $p$ is not prime, it is also possible to condense a subset of the magnetic fluxes.  This confines only a subset of electric flux loops (corresponding to the charges that have nontrivial Berry phase with the condensed vortex lines), and leaves the remaining electric fluxes and their corresponding charges deconfined. 
Thus  there are also condensation transitions between the $\integ_{p}^\text{Potts}$ phase and a $\integ_{m}^\text{Potts}$ phase, where $m$ divides $p$.  An example of the general structure of these phase diagrams is shown for $p=4$ in Fig. \ref{Z4Fig}.  In all of these transitions, there is a {\it reduction} in the ground state degeneracy, and topological entanglement entropy, as the system enters the condensed phase.  

In contrast, in the uncondensed phase of the \AWW models, there are no deconfined excitations in the bulk.  The ground state is nonetheless a ``string-net", with loops (or nets, for $p>2$) of non-trivial ``electric flux" (i.e. non-zero spin, in the language of Sect.s \ref{PottsSec} and \ref{AWWSec}).  (Here we use ``electric" and ``magnetic" flux by analogy with the corresponding objects in the $\integ_{p}^\text{Potts}$ gauge theory, although for the \AWW models the analogy is not exact).  Because the ground state contains extended loops (or nets), here too we can add a perturbation that condenses loops of ``magnetic flux" (i.e. plaquette defects), driving a phase transition that confines ``electric flux" (i.e. edges carrying certain spin labels).  As for the $\integ_{p}^\text{Potts}$ case, simultaneously condensing all magnetic flux loops engenders a transition to a trivial phase, in which all electric flux lines are confined.  If $m$ divides $p$, however, we may once again condense a subset of the possible magnetic fluxes, leaving $m-1$ deconfined types of electric flux loops.  Surprisingly, many of the latter condensed phases have bulk topological order and deconfined bulk excitations: the topological ground state degeneracy {\it grows} as the system is driven into the condensed phase.

\begin{figure}
 \includegraphics[width=0.975\linewidth]{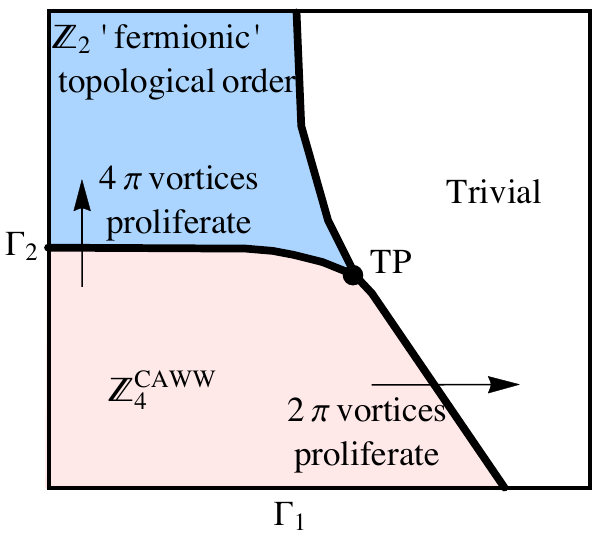}
 \caption{(Color online) A sketch of the phase diagram for the $p=4$ \AWW model, as a function of the transverse field strengths $\Gamma_1$, ($2 \pi$ vortex loops) and $\Gamma_2$ ($4 \pi$ vortex loops), where charge is quantized in multiples of $1/4$ (mod 1).  For small $\Gamma_1, \Gamma_2$, there are extended loops of electric flux, as in the solvable $p=4$ \AWW model.  This region is separated by first-order phase transitions from both the trivial region at large $\Gamma_1$ and a phase described by a  $\integ_{2}$ gauge theory with fermionic matter sources at large $\Gamma_2$.  
 As we have shown, the locations of the phase boundaries are identical to those of Fig. \ref{Z4Fig}. }
\label{WW4Fig}
 \end{figure}

As we discussed in Sect. \ref{AWWSec}, in general we cannot rule out the possibility that the uncondensed phase of these \AWW models is in fact connected to the trivial phase in which all electric fluxes are confined, once we allow perturbations that introduce vertex violations into the ground state.   (We note, however, that there are Abelian Walker-Wang models similar to those discussed here that are known to be symmetry-protected phases.\cite{WWSPT})  Interestingly, when a {\it subset} of the possible vortex loops is condensed, there are transitions across which the topological order changes; these are necessarily transitions between distinct phases.  Fig. \ref{WW4Fig} sketches the possible forms of the phase diagram for $p=4$, as a function of the two transverse fields $\Gamma_1, \Gamma_2$ which create vortex loops of magnetic flux $2 \pi$ and $4 \pi$, respectively.  For small $\Gamma_1, \Gamma_2$, there is a \AWW region.  This region is separated from the regions at large $\Gamma_1, \Gamma_2$ by a first-order phase transition.  For large $\Gamma_2$, the system is  in a  $\integ_{2}^\text{Potts}$ phase (with the slight twist that the charges are fermions rather than bosons).  Since this phase is topologically ordered, the phase boundary must persist in the presence of arbitrary perturbations to the Hamiltonian.  For large $\Gamma_1$, the system is in a trivial phase with all electric fluxes confined.  
The arguments of Appendix \ref{PhaseTranApp} ensure that the phase boundaries match those of the analogous \zp Potts-type model shown in Fig. \ref{Z4Fig}.  
However, since neither topological order nor symmetry distinguish this phase from the uncondensed \AWW phase, the phase boundary need not persist when we allow arbitrary perturbations (which do not commute with the vertex term $\oQ_V$) to the Hamiltonian.   

We have also briefly discussed transitions in more complicated \AWW models, for which the surfaces admit states with non-Abelian topological order.   In certain cases we are able to identify the nature of the phase transition here too (which is again first order) by mapping the transitions onto transitions in the $\integ_{p}^\text{Potts}$ gauge theories.  We have also outlined a completely general procedure for deducing the topological order of any condensed phase, and identified transitions from \AWW models with the surface topological order of $SU(2)_k$ Chern-Simons theories to phases with bulk topological order.  Thus the fact that a condensation transition that partially confines the allowed flux (or spin) labels can {\it deconfine} some of the remaining charges in the bulk is not unique to the Abelian case.

\begin{figure*}
 \includegraphics[width=0.9\linewidth]{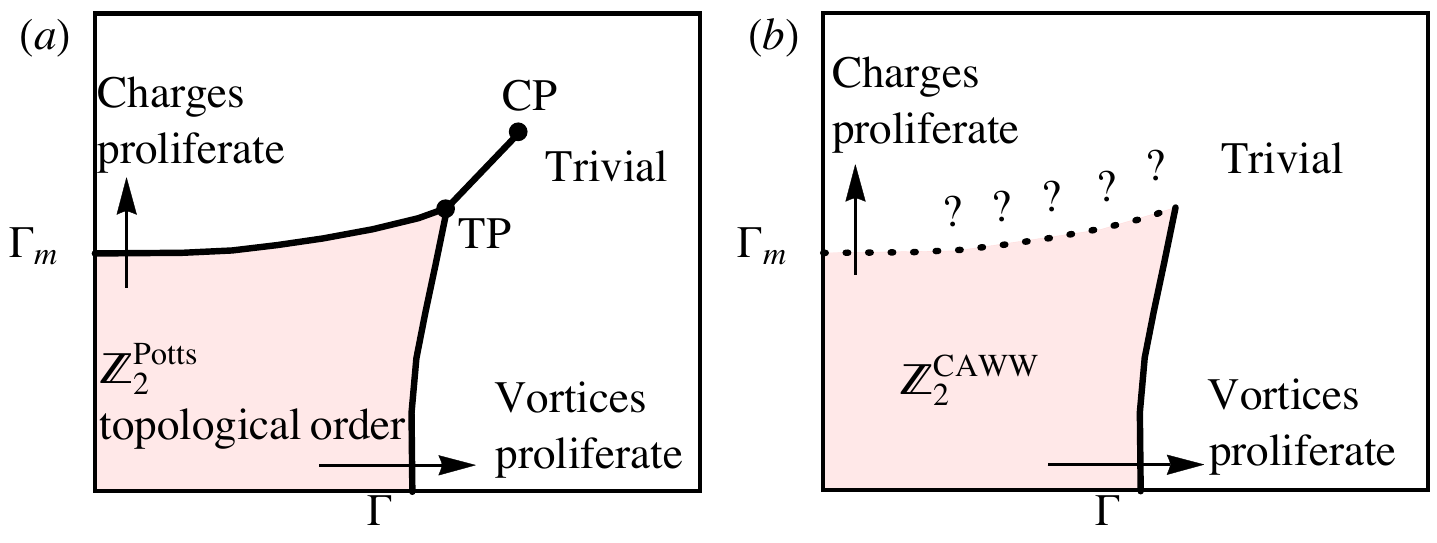}
 \caption{(Color online): A sketch of the phase diagram both $\integ_{2}^\text{Potts}$ (a) and $p=2$ \AWW (b) models, with an addional transvere field term $\Gamma_m$ that can be used to drive a Higgs transition.  For the $\integ_{2}^\text{Potts}$ model, there are two phases: the topologically ordered $\integ_{2}^\text{Potts}$ phase, and the trivial phase which is obtained by condensing either charges or vortex loops.  These are separated by a first-order transition (solid line) along the $\Gamma$ axis, and a second-order (dotted line) 4D Ising transition along the $\Gamma_m$ axis\cite{CreutzZ2,Ising4DMC}.  A first-order line emerges at the point where these two meet, terminating at a second-order critical point.  The phase boundaries sketched here are based on the numerical work of Ref. \onlinecite{CreutzZ2}.  For the \AWW model, we do not expect the phase diagram to match that of the $\integ_{2}^\text{Potts}$ model once $\Gamma_m>0$, so that the loci and nature of the phase transitions are largely currently unknown.  In particular, though the Higgs phase clearly exists (and has the same ground state, in the limit $\Gamma_m \rightarrow \infty$, as for the $\integ_{2}^\text{Potts}$ case), we do not know at present whether there is a phase transition along the $\Gamma_m$ axis, or indeed to what values of $\Gamma_m$ the transition along the $\Gamma$ axis persists.  
 }
\label{AWWMFig}
 \end{figure*}

Returning to our $\integ_{p}^\text{Potts}$ gauge theories, we note that in these systems there are generically {\it two} types of condensation transitions: the first-order confining transition, in which vortex loops or monopoles proliferate, and the Higgs transition, in which charges condense.  (The Higgs transition is second-order for $p=2$, and first-order for $p\geq 3$\cite{BaigPhysLettB207}).  It is known\cite{FradkinShenker} that the Higgs and confined phases are not distinct (provided we condense the fundamental charge or fundamental flux).  The qualitative form of the phase diagram in this more general case is shown in Fig. \ref{AWWMFig}, for the example $p=2$.  

It is thus natural to ask about the analogue of the Higgs transition in the \AWW models.  
To drive such a transition, we must introduce an additional transverse field term (of the form $ \Gamma_m \oW_e$) which raises the value of the spin on each edge $e$.  Unlike the case of  $\integ_{p}^\text{Potts}$ gauge theories, however, such an operator fails to commute not only with the vertex terms at each end of the edge, but also with the  plaquette terms on some of the surrounding plaquettes.  Thus we can enter the Higgs phase only by making $\Gamma_m$ large compared with {\it both} the coefficient of the vertex term {\it and} the coefficient of the plaquette term.  Because of this complication, the Higgs transition in the \AWW models does not map onto that of the  $\integ_{p}^\text{Potts}$ gauge theories in a straightforward way.  Indeed, it is not obvious that there is a phase transition at all, given that there is no clear indicator that the \AWW model represents a distinct phase.  What is clear, however, is that for sufficiently large $\Gamma_m$ the model is again in a trivial phase, with the spins on each edge diagonal in the $\oW_e$ basis in the limit $\Gamma_m \rightarrow \infty$.  
Further, as this trivial phase (together with the trivial confined phase) is identical to its counterpart in the  $\integ_{p}^\text{Potts}$ gauge theory, the arguments of Ref. \onlinecite{FradkinShenker} ensure that the $\Gamma \ra \infty$ and $\Gamma_m \ra \infty$ phases are connected.  A qualitative sketch of the resulting phase diagram is shown in Fig. \ref{AWWMFig}.

Of course, in the  $\integ_{p}^\text{Potts}$ gauge theory, we may also condense charges other than the fundamental one, to obtain Higgs phases with non-trivial ( $\integ_{m}^\text{Potts}$) topological order.  In the \AWW models, such transitions appear not to produce new topologically ordered phases, however.  

In summary, this work has detailed just some of the many possible phase transitions that can occur in the Walker-Wang models.  Though in the limits that we are able to describe, the transitions themselves are not exotic, the phase diagram has some very surprising features, including the possibility of producing topological order from a phase which cannot be distinguished, by either topological order or symmetry, from the trivial phase.  We hope that future numerical or analytical work will be able to shed light on this rather surprising phenomenon.  

{\bf Acknowledgements:}
CVK acknowledges the financial support of the EPSRC. SHS acknowledges funding from EPSRC grants EP/I032487/1 and EP/I031014/1.

\appendix

\section{General form of the plaquette projector for Confined Abelian Walker-Wang models} \label{ThetaApp}

\begin{figure}
\begin{center}
\includegraphics[width=.975\linewidth]{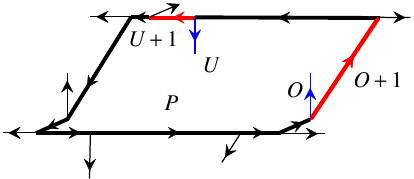}
                \caption{(Color online): Orientations of edges used in the definition of the Walker-Wang Plaquette operator, shown for the 3 different types of plaquettes on the point-split cubic lattice (see main text).  Edges in $\partial P$ are shown in bold; edges in $*P$ are not bold.  The definition of the plaquette operator requires 2 special edges in $*P$, shown here in blue and indicated with the letters O and U.  For $p$ even we also require two special edges O$+1$, U$+1$, which are shown here in red. }
                \label{PlaqFig}
\end{center}                
\end{figure}

Here we give a general expression for the phase operator $\hat{\Theta}_{P,m}$ in Eq. (\ref{BpWW}).  As discussed in the main text, $\hat{\Theta}_{P,m}$ is diagonal in the spin basis, and has eigenvalues $e^{ i \pi n m /p}$, $0 \leq n < 2 p-1$.  Thus $\hat{\Theta}_{P,m}$ can be expressed in terms of a product of the operators $\oS_{e, \pm}$ acting on certain edges proximate to the plaquette $P$.  
 In order to express $\hat{\Theta}_{P,m}$ in this way, however, for $p>2$ we must excercise some care about our choice of $\oS_{e,+}$ versus $\oS_{e,-}$.  To avoid confusion, our choice is indicated in Fig. \ref{PlaqFig}: an arrow pointing in the $\hat{x}, \hat{y}$, or $\hat{z}$ direction on edge $e$ indicates that we use $\oS_{e,+}$ and $\oW_{e,+}$; an arrow along  $-\hat{x}, -\hat{y}$, or $-\hat{z}$ on $e$ indicates that we use $\oS_{e,-}$  and $\oW_{e,-}$.  
 
 Let us begin by defining operators that measure the spin (rather than its exponential) on each edge:
 \ba
 \os_{e, +} |s_e \rangle &=& s_e |s_e \rangle \n
  \os_{e, -} |s_e \rangle &=&(p -  s_e) |s_e \rangle
 \ea
 where $s_e \in \{ 0,1, ... p-1 \}$.  In terms of the operators $\oS$ introduced in Eq. (\ref{Sops}), we have
 \be
 \oS_{e, \pm} = \text{Exp} \left [  i \ ( 2 \pi /p)  \ \os_{e, \pm} \right ]
 \ee

 \subsection{$p$ odd}
 
For technical reasons (which we discuss shorttly), the definition of the operator $\hat{\Theta}_{P,m} $ depends on whether $p$ is even or odd.  
To define $\hat{\Theta}_{P, m}$, we identify two special ``crossed" edges in $*P$.  (Recall that $*P$ is the set of all edges connected to, but not bordering, the plaquette $P$  
 (these are the non-bold edges in Fig. \ref{PlaqFig}).)  Each plaquette contains one ``O" and one ``U" edge, indicated in blue in the Figure.  

For $p$ odd, we take:
\be \label{theta1}
\hat{\Theta}_{P, m} = \text{Exp} \left[ i  \frac{ \pi \ m (p-1)}{p} \left( \os_{\text{O}, \nu_\text{O}} -  \os_{ \text{U}, \nu_\text{U} } \right )  \right ]
\ee
 where the subscripts $ \text{U}$ ($ \text{O}$ ) denote the  U (O) edges, with orientations chosen as shown in the Figure.  

 \subsection{$p$ even}
 
For even $p$, the choice of $\hat{\Theta}_{P, m}$ in Eq. (\ref{theta1}) does not lead to a well-defined Walker-Wang model.  
This is because changing the spin on an O or U edge by $p$ leads to an overall phase of 
$\text{Exp} \left[  i  \pi \ m (p-1) \right ] = (-1)^m$.  This is inconsistent with the notion that the spin state on each edge is defined only modulo $p$.

Instead, the phase operator that we use for even $p$ can be expressed as:
\be \label{ThetaDef}
\hat{\Theta}_{P,m} =  \left( -1\right)^m \text{Exp} \left [ i \frac{ \pi  \ m}{p}  \sum_{e_i \in *P}  \os_{e_i, \nu_i} \right ] \tilde{\theta}_{P, m}
\ee
where the values of $\nu_i = \pm 1$ are indicated in Fig. \ref{PlaqFig}, as explained above, and $ \tilde{\theta}_{P, m}$ is defined below.  Note that $\sum_{e_i \in *P}  \os_{e_i, \nu_i}$ is always a multiple of $p$, provided that the vertex condition is always satisfied. (When the vertex condition is not satisfied, the expression for $\hat{\Theta}_{P,m}$ is more complicated, but can be deduced from the form of the Hamiltonian given in Ref. \onlinecite{WW}.)  Thus for $m$ even, the first two terms in the product (\ref{ThetaDef}) are always $1$.  For $m$ odd, their combination is negative if the net spin of all edges leaving the plaquette (with the orientations show in Fig. \ref{PlaqFig}) is an even multiple of $p$, and positive if it is an odd multiple of $p$.   
 
 With the same definitions of O and U edges as used above, we have:
 \ba
 \tilde{\theta}_{P, m} &=& \text{Exp} \left[ i \frac{ \pi \ m }{p} \left( \os_{\text{O}, \nu_\text{O}} -  \os_{ \text{U}, \nu_\text{U} } \right )  \right ] \times \\
&& \left( \text{sign} ( \os_{\text{U}+1, \nu_{\text{U}+1} } -m ) \right)^{\os_{\text{U}, \nu_{\text{U}}}}  \times \n 
&& \left( \text{sign} ( \os_{\text{O}+1, \nu_{\text{O}+1}} -m ) \right)^{\os_{\text{O}, \nu_{\text{O}}}} \nonumber 
 \ea
 where  $\text{U}+1$ ($\text{O}+1$) are edges in $\partial P$ that are adjacent to the O and U edges,  as shown in the Figure.   Note that here we take sign(0) = 1.  
 
 Evidently, since $p-1$ and $p$ are coprime, in either case the possible eigenvalues of $\hat{\Theta}_{P,m}$
are $e^{ i \pi n m /p}, 0 \leq n < 2p$.

 \bigskip
 
Readers familiar with TQFTs may wish to note that for $p$ odd, we use the category $SU(p)_1$ (or $\mathbb{Z}^{(N/2-1/2)}_N$, in the notation of Ref. \onlinecite{BondersonThesis}), while for $p$ even, we use $U(1)_{p/2}$ (or $\mathbb{Z}^{(1/2)}_N$).\footnote{Detailed information about these two categories can be found, for example, in Ref. \onlinecite{BondersonThesis}.}   For odd $p$ the $6j$ symbols are all unity, and $\hat{\Theta}_{P,m}$ is defined purely by the $R$ matrices.  For even $p$ the $6j$ symbols may be negative; the term $\text{Exp} \left [ i \frac{ \pi m}{p}  \sum_{e_i \in *P}  \os_{e_i, \nu_i} \right ]$ in Eq. (\ref{ThetaDef}) gives the net effect of the $6j$ symbols required to fuse the spin $m$ with the spins on the edges of $P$, with orientations given in Fig. \ref{PlaqFig}.  (This term is also present in the  2D Levin-Wen\cite{LW} models, on which Walker and Wang's construction is heavily based.)  The term $\tilde{\theta}_{P, m}$ is unique to 3D, %(where the $R$ matrices enter into the definition of $B_P$)
 and accounts for crossings of the string labelled $m$ with certain edges in $*P$.  (This is explained in more detail in Ref. \onlinecite{WWUs}).  
The choice of ``O" and ``U" edges depends on the projection (or angle of view) used to determine these crossings.  

\subsection{Spectrum of the plaquette term}

Let us begin by deducing a few key properties of $\hat{B}_{P, m} \equiv \left( \hat{\Phi}_{P}\right)^m \hat{\Theta}_{P, m}$ necessary for understanding the plaquette projector.  First, we prove that
\be
\hat{B}_{P, m} =  \left( \hat{B}_{P, 1} \right )^m
\ee
For odd $p$, this is immediate since $\left[ \hat{\Phi}_P, \hat{\Theta}_{P, m} \right ] =0$ ($\hat{\Theta}_{P,m}$  acts only on edges that are not raised by $\hat{\Phi}_P$), and from Eq. (\ref{theta1}) it is clear that $  \hat{\Theta}_{P,m}=  \left( \hat{\Theta}_{P,m}\right)^m$.  
For even $p$,  it is a question of verifying that the two sign terms are indeed shifted appropriately.  (All other terms commute).  To see this, observe that:
\begin{widetext}
\be
\left( \text{sign} ( \os_{\text{U}+1, \nu_{\text{U}+1} } -m ) \right) \hat{\Phi}_P |s_{\text{U}+1} \rangle
 =\hat{\Phi}_P \ \text{sign} \left(  \os_{\text{U}+1, \nu_{\text{U}+1} } -\text{mod }( m +1, p)   \right) \ \text{sign} \left(  \os_{\text{U}+1, \nu_{\text{U}+1} } -1   \right)  |s_{\text{U}+1} \rangle 
\ee
This holds because (with the choice of orientations given in Fig. \ref{PlaqFig}) $\hat{\Phi}_P$ lowers the eigenvalue of $ \os_{\text{U}+1, \nu_{\text{U}+1} }$ by $1$(mod $p$); to compensate for this we have raised $m$ by $1$ (mod $p$) on the right-hand side.   %The extra factor of sign$ \left(  \os_{\text{U}+1, \nu_{\text{U}+1} } -1   \right) $ ) on the right-hand side accounts for the fact that the label $0$ is lowered to $p-1$.   
(Evidently, this identity also holds if we replace U by O).  From this, we deduce that: 
\begin{align}
&\left[ \hat{\Phi}_P   \left( \text{sign} ( \os_{\text{U}+1, \nu_{\text{U}+1} } -1 ) \right)^{\os_{\text{U}, \nu_{\text{U}}}}  \left( \text{sign} ( \os_{\text{O}+1, \nu_{\text{O}+1}} -1 ) \right)^{\os_{\text{O}, \nu_{\text{O}}}} \right]^m \n
& = \left( \hat{\Phi}_P\right)^m   \left( \text{sign} ( \os_{\text{U}+1, \nu_{\text{U}+1} } -m ) \right)^{\os_{\text{U}, \nu_{\text{U}}}}  \left( \text{sign} ( \os_{\text{O}+1, \nu_{\text{O}+1}} -m ) \right)^{\os_{\text{O}, \nu_{\text{O}}}}
\end{align}
as required.  
\end{widetext}

Second, it follows that
\be \label{BPPowerEq}
\left( \hat{B}_{P, 1} \right )^p = \left( \hat{\Phi}_{P}\right)^p \hat{\Theta}_{P, p}  = 1
\ee
This is because $\left( \hat{\Phi}_{P}\right)^p = \hat{\Theta}_{P,p} = 1$.  For $p$ odd, the latter follows immediately from Eq. (\ref{theta1}), as $p-1$ is even.  For $p$ even, we have that $\text{ Exp } i \pi \left[ \sum_{e_i \in *P}  \os_{e_i, \nu_i} \right ] =1 $, and
%\ba
%&\left[ \left( \hat{\Phi}_{P}\right)^m  \left( -1\right)^m \text{Exp} \left [ i \frac{ \pi  \ m}{p}  \sum_{e_i \in *P}  \os_{e_i, \nu_i} \right ] \tilde{\theta}_{P, m}  \right]^p \n
%=& 
%\left[ \left( \hat{\Phi}_{P}\right)^m  \tilde{\theta}_{P, m}  \right]^p  \text{Exp} \left [ i  \pi \ m \sum_{e_i \in *P}  \os_{e_i, \nu_i} \right ] \n
% =& 
%\left[ \left( \hat{\Phi}_{P}\right)^m  \tilde{\theta}_{P, m}  \right]^p
%\ea
%where we have used $\left( -1 \right )^{m p} =1$, since $p$ is even, and noted that $\text{Exp} \left [ i  \frac{\pi \ m}{p} \sum_{e_i \in *P}  \os_{e_i, \nu_i} \right ]$ commutes with $\hat{\Phi}_P$, since it acts only on edges in $*P$.  The last equality holds because $\sum_{e_i \in *P}  \os_{e_i, \nu_i}$ is a multiple of $p$.  
 $ \tilde{\theta}_{P, p} = e^{ i \pi  \left( \os_{\text{O}, \nu_\text{O}} -  \os_{ \text{U}, \nu_\text{U} } \right )  }  \left( \text{sign} ( \os_{\text{U}+1, \nu_{\text{U}+1} }  ) \right)^{\os_{\text{U}, \nu_{\text{U}}}}  \times \n  \left( \text{sign} ( \os_{\text{O}+1, \nu_{\text{O}+1}}  ) \right)^{\os_{\text{O}, \nu_{\text{O}}}}= 1$.

\subsection{Confinement and deconfinement in Abelian Walker-Wang models}

We next show in more detail how the presence of $\hat{\Theta}$ in the plaquette term leads to confinement of all excitations in the bulk -- and why eliminating certain labels from the effective action can produce deconfined excitations.  

There are two types of excitations in these models: vortex loops (which always incur an energy cost per unit length, for $\Gamma=0$) and vertex defects, which correspond to deconfined charges in the Potts gauge theories.  A pair of vertex defects at vertices $v_1$ and $v_2$ is created by raising the spin along a series $C_{12}$ of edges connecting $v_1$ to $v_2$.  We will call the operator that does this $V^\dag_{C_{12}}$.  The excitations are deconfined if and only if there exists a raising operator that commutes with $\hat{B}_P$ for all $P$ (except possibly those plaquettes of which $v_1$ and $v_2$ are vertices).   This is most easily seen at the exactly solvable point $\Gamma=0$, where the eigenvalue of $\hat{B}_P$ is conserved.  Since the spectrum of $\hat{B}_P$ is discrete, such strings either cost no energy (if they commute with $\hat{B}_P$) or incur a fixed energy penalty per unit length (if they do not). 

Consider the effect of $V^\dag_{C_{12}}$ for a trajectory $C_{12}$ that contains, say, the O edge (but not the $U$ edge) of some plaquette $P$.  If $V^\dag_{C_{12}; r}$ simply raises the spins on all edges by $r$, then we have:
\be
\hat{\Theta}_{P,m} V^\dag_{C_{12}; r} = \begin{cases} e^{ - i \pi m (p-1)/p \ r} V^\dag_{C_{12};r} \hat{\Theta}_{P,m} & p \text{  odd}  \\
\pm e^{ - i \pi m /p \ r} V^\dag_{C_{12};r} \hat{\Theta}_{P,m} & p \text{  even}
\end{cases}
\ee 
The commutators are unchanged if $C_{12}$ contains the U edge (but not the O edge), since the O and U edges by definition have opposite orientations.   

We could also consider operators that simultaneously raise the spin on edges in $C_{12}$, and multiply the result by a phase depending on the spins on edges in $*C_{12}$ (i.e. on the set of edges connected to, but not on, the path $C_{12}$).  For example, if our operator acts on the O edge of a plaquette $P$, we can eliminate the phase difference between $\hat{B}_{P,m} V^\dag_{C_{12}}$ and  $V^\dag_{C_{12}}\hat{B}_{P,m}$ by adding to $V^\dag$ a phase of the form $\text{Exp} \left[ i \frac{ \pi \  (p-1)}{p} \os_{O+1, \nu_{O+1}} \os_{O, \nu_O} \right ] $ for $p$ odd (or its equivalent for $p$ even).  This will give an operator that commutes with $\hat{B}_P$, but {\it fails} to commute with the plaquette term on at least one of the neighboring plaquettes (with which it commuted previously).  Thus in general, 
it is not possible to create an operator $V_{C_{12}}$ that does not incur a finite energy cost per unit length.  

Next, let us consider what happens when we restrict ourselves to a subset of the possible values of $m$.  Suppose, for example, that both $m$ and $r$ must be multiples of $p/s$, where $s|p$  --i.e. $m =  a p/s, r = b p/s$.  With this restriction, the possible phases appearing in the commutator of $\hat{\Theta}_{P,m}$ and $V^\dag_{C_{12}; r}$ are integer multiples of $ i \pi b  p/s^2 $.   If $b p/s^2$ is not an integer, then the situation is as described above: if $C_{12}$ contains both the O and U edges of $P$, $\left[ V^\dag_{C_{12}; b p/s}, \hat{\Theta}_{P,p/s} \right ] \neq 0$,  and there is no way to adjust $V^\dag$ that avoids paying a finite energy cost per unit length.  
 Essentially the same reasoning can be used to show that there is no operators $\oW_C$ that raises all spins on the non-contractible curve $C$ by $r = b p/s$ and commute with the Hamiltonian.  
 
Conversely, if $b p/s^2$ is an integer,  then there is an operator $V^\dag_{C_{12}; r}$ that commutes with $\hat{B}_{P, a p/s}$ for every integer $a$ and plaquette $P$.   If, for $C_{12}$ containing either the O or the U edge of $P$ (but not both),   $\left[ V^\dag_{C_{12};r}, \hat{\Theta}_{P,p/s} \right ] = 0$, then $V^\dag_{C_{12}; r}$ simply raises all spins along $C_{12}$ by $r$, as for the Potts gauge theory.  If $\left \{ V^\dag_{C_{12};r}, \hat{\Theta}_{P,p/s} \right \} = 0$, then $V^\dag_{C_{12}; r}$ also contains phases depending on the spin labels in $*C_{12}$; nonetheless it is possible to assign these such that $\left[ V^\dag_{C_{12}; r}, B_{P, p/s} \right ] =0$.  A simple way to see that there must be deconfined particles in this case is to note that $V^\dag_{C_{12}}$ commutes with $\hat{B}_{P, a p /s}$ whenever $C_{12}$ contains both the O and the U edge of $P$, allowing us to separate charges along certain directions without any energetic penalty.  By adding extra phases to $V^\dag_{C_{12}}$, we can allow separations in any direction.

To summarize, there is one deconfined particle species for each $b$ such that $e^{ i \pi b  p/s^2} = \pm 1$, as claimed in the main text.  For each deconfined particle species, there is a corresponding operator $\oW_C$ (for each non-contractible curve $C$) that commutes with the Hamiltonian, and can be used to map between topologically distinct ground-state sectors.

\section{Counting spin configurations in general Potts gauge transitions} \label{CountApp} 

Here we verify that the condensation transition involving $\hat{\bf{h}}_e^{(m)}$ in a $\mathbb{Z}_{p}^\text{Potts}$ model can be mapped exactly onto a transition involving $\hat{\bf{h}}_e^{(1)}$ in a  $\mathbb{Z}_{p/m}^\text{Potts}$ model.

Our objective is to show that, for given $\Gamma, \lambda$, the expectation value of any operator involving the flux excitations present in the ground state (i.e. those created by $\hat{\bf{h}}_e^{(m)}$) is the same as the expectation of the analogous operator in the $\mathbb{Z}_{p/m}^\text{Potts}$ model.  
(Since the excitations  that condense in both cases are loops of flux $2 \pi n m/p$, the analogous operator is simply the operator creating the same configuration of flux loops).  

Intuitively, this follows from the fact that each insertion of $\left( \oS_e^{m } \right)^n$ creates one flux loop (with flux $\phi_P =  2 \pi (m/p) n$) encircling the edge $e$, and that the rules for combining the flux loops are the same in both models.  This ensures that the number of ways of obtaining a given configuration of flux loops within perturbation theory is the same in both cases.

Let us see how this arises at the level of configurations.   Since $ \left ( \hat{\Phi}_P\right )^{ n p/m} $ commutes with all operators in the Hamiltonian, energetically there is no difference between an edge of spin $s$ and an edge of spin $p/m + s$ (mod $p$).  In other words,
the many-to-one mapping between configurations of the $\mathbb{Z}_{p}^\text{Potts}$ model and the $\mathbb{Z}_{p/m}^\text{Potts}$ model
\be
s_e + n p /m \text{  ( mod } p)  \rightarrow s_e
\ee
preserves the energy of any spin configuration.  In particular, it does not affect the configuration of those vortex loops created by $\hat{\bf{h}}_e^{(m)}$, as $\left[ \hat{\bf{h}}_e^{(m)} , \left( \oW_{e, \pm} \right)^{p/m}\right ] =0$.  Hence the expectation value of any operator involving the vortex loops (i.e. any operator that is related to the degrees of freedom involved in the phase transition) is unaffected by this mapping.  It follows that the phase transition must be identical in both systems.  

Because the mapping is many-to-one, it fails to capture the physics of the degrees of freedom that are not involved in the phase transition.  Indeed, the mapping sends $m$ distinct spins of the $\mathbb{Z}_{p}^\text{Potts}$ model to a single spin in the $\mathbb{Z}_{p/m}^\text{Potts}$ model.  Hence there are $m$ deconfined spin variables in the condensed phase, as opposed to a single ($s_e =0$) deconfined spin variable in the condensed phase for the $\mathbb{Z}_{p/m}^\text{Potts}$ case.

\section{Equivalence of Phase transitions in Potts and  Walker-Wang models} \label{PhaseTranApp}

Here we give detailed arguments as to why certain phase transitions in Walker-Wang models must be of the same type as transitions known in the $\mathbb{Z}_{p}^\text{Potts}$ models.  

\subsection{Abelian Walker-Wang models}

We begin with the Abelian Walker-Wang models.  Here all possible phase transitions can be mapped onto transitions in the $\mathbb{Z}_{p}^\text{Potts}$ models.   In practice,  the arguments presented here do not depend on the relative coefficients of the different powers of $\hat{B}_{P,1}$ and $\oS_{e,+}$ in the definitions of $\hat{B}_P$ and $\oh_e$, and can equally be applied to variants of the \AWW models, such as the \zp  lattice gauge theory of Sect. \ref{ZpClockSect}, or the $\Gamma_1 - \Gamma_2$ models discussed in the Conclusion.  

Specifically, 
 we will show that the expectation value of any operator composed of an arbitrary sum of products of 
 flux measurements and spin measurements is identical in the two models, for any value of $\Gamma/ \lambda$.  In the \AWW models, flux measurements are carried out by the operator
 $\hat{B}_{P,1} \equiv \left( \hat{\Phi}_P \hat{\Theta}_{P,1}  \right )$, whose eigenvalues we identified with the flux through a given plaquette.  In the $\mathbb{Z}_{p}^\text{Potts}$ model, a flux measurement is given by $\hat{B}_{P,1}  = \hat{\Phi}_P$.  The two flux operators differ by a set of spin measurements, carried out by $\oS_e$ in both models.  Hence the set of operators that we allow is identical in both models; however, it is convenient to identify our ``flux" operator as the object that commutes with the Hamiltonian for $\Gamma=0$.

First, we note that since $\left[ \hat{\Theta}_{P,m}, \oh_e \right ] =0$, we have
\be\label{EdgeJunk}
\left( \hat{\Phi}_P \hat{\Theta}_{P,1}  \right )\oS_{e, \nu_e} =  \begin{cases} e^{ - 2 \pi i /p } \oS_{e, \nu_e} \left( \hat{\Phi}_P \hat{\Theta}_{P,1}  \right ) & \ \ e \in \partial P \\
 \oS_{e, \nu_e} \left( \hat{\Phi}_P \hat{\Theta}_{P,1}  \right ) & \ \text{ otherwise} 
\end{cases}
\ee
 In the  $\mathbb{Z}_{p}^\text{Potts}$ models, the identical relationship holds, with $\left( \hat{\Phi}_P \hat{\Theta}_{P,1}  \right)$ replaced by $\hat{\Phi}_P$.
In either model, therefore, any product $\hat{\mathcal{O}}$ of $\hat{B}_{P_i,1}$  and $\oS_{e_k, \nu_{e_k}}$ can be expressed in the form:
\be \label{Oform}
\hat{\mathcal{O}}  = e^{ -i 2 \pi q/p} \prod_i \left( \oS_{e_i, \nu_i} \right)^{n_i} \prod_j  \left( \hat{B}_{P_j, 1} \right)^{n_j} 
\ee
where $q$ is the same integer for the $\mathbb{Z}_{p}^\text{Potts}$ and \AWW cases.  

Second, observe that in both models, the exact ground state $|\Psi_0 \rangle$ for any value of $\Gamma/ \lambda$ can be expressed in the form:
\be \label{0Form}
|\Psi_0 \rangle = \sum_\alpha a_\alpha^{(\Gamma/ \lambda)} \prod_{k \in \{ \mathcal{C}_\alpha \} }  \oh_{e_k} |\Psi_0 \rangle_{\Gamma=0}
\ee
where $\{ \mathcal{C}_\alpha \}$ is an arbitrary collection of edges in the lattice, and $a_\alpha$ are arbitrary coefficients.  This is true everywhere in the phase diagram, though the coefficients $a_\alpha$ can be evaluated perturbatively only in the uncondensed phase.  Thus the expectation value of an operator  $\hat{\mathcal{O}}$ can always be evaluated via:
\begin{widetext}
\be
\langle \hat{\mathcal{O}} \rangle_{\Gamma, \lambda} =\langle \Psi_0|_{\Gamma=0}   \sum_{\alpha, \beta}  \left( a_\alpha^{(\Gamma/ \lambda)} \right)^*   a_\beta^{(\Gamma/ \lambda)}  \left( \prod_{k \in \{ \mathcal{C}_\alpha  \} }  \oh_{e_k} \right )  \hat{\mathcal{O}} \left( \prod_{j \in \{ \mathcal{C}_\beta  \} }  \oh_{e_j} \right )   |\Psi_0 \rangle_{\Gamma=0}
\ee

If $\hat{\mathcal{O}}$ is a sum of products of $\hat{B}_{P_1}$ and $\oS_e$, we may use Eq. (\ref{EdgeJunk}) repeatedly to move all $\hat{B}_{P_1}$ operators to the right,  arriving at the form:
\be
\langle \hat{\mathcal{O}} \rangle_{\Gamma, \lambda} =\langle \Psi_0|_{\Gamma=0}   \sum_{\alpha, \beta}  \left( a_\alpha^{(\Gamma/ \lambda)} \right)^*   a_\beta^{(\Gamma/ \lambda)}  \left( \prod_{k \in \{ \mathcal{C}_\alpha  \} }  \oh_{e_k} \right ) \left( \prod_{j \in \{ \mathcal{C}_\beta  \} }  \oh_{e_j} \right ) \sum_\gamma  c_\gamma \tilde{\hat{\mathcal{O}}}_\gamma  |\Psi_0 \rangle_{\Gamma=0}
\ee
where $ \tilde{\hat{\mathcal{O}}}_\gamma $ is of the form (\ref{Oform}), and the coefficients $c_\gamma$ are necessarily the same in both models as they follow directly from Eq. (\ref{EdgeJunk}).
Since $\hat{B}_{P, 1}|\Psi_0 \rangle_{\Gamma=0} = |\Psi_0 \rangle_{\Gamma=0}$, we have 
\be \label{OExt}
\langle \hat{\mathcal{O}} \rangle_{\Gamma, \lambda} =\sum_{\alpha, \beta, \gamma } \left(  a_\alpha^{(\Gamma/ \lambda)} \right )^*  a_\beta^{(\Gamma/ \lambda)} c_\gamma  \langle \Psi_0|_{\Gamma=0}   \left( \prod_{k \in \{ \mathcal{C}_\alpha  \} }  \oh_{e_k} \right ) \left( \prod_{j \in \{ \mathcal{C}_\beta  \} }  \oh_{e_j} \right )  \prod_{i } \left( \oS_{e_i, \nu_i} \right)^{n_i}   |\Psi_0 \rangle_{\Gamma=0}
\ee
\end{widetext}
 In other words, we can express $ \langle \hat{\mathcal{O}} \rangle$ as a sum of expectations of operators that are diagonal in the spin eigenbasis, with coefficients that are completely determined by $\{ a_\alpha^{(\Gamma/ \lambda)}  \}$ and the relation (\ref{EdgeJunk}).  

To complete the proof, we must show two things.  First, that if $\hat{\mathcal{O}}_s$ is an operator diagonal in the spin basis, 
$
\langle \Psi_0|_{\Gamma=0}\hat{\mathcal{O}}_s   |\Psi_0 \rangle_{\Gamma=0}
$
is the same for both models; and second, that the coefficients $a_\alpha^{(\Gamma/\lambda)}$ in Eq. (\ref{0Form}) are the same for the \AWW  and $\mathbb{Z}_{p}^\text{Potts}$ ground states.

 For the first item, we observe that the $\Gamma =0$ ground state can be constructed from the trivial state $|0 \rangle \equiv \prod_e | s_e = 0 \rangle $, via
\be \label{GSMake}
|\Psi_0 \rangle_{\Gamma =0} = \prod_P \hat{B}_P  |0 \rangle
\ee
Since $\hat{B}_P^2 = \hat{B}_P$ (or more generally, $\hat{B}_{P,1} \hat{B}_P = \hat{B}_P$, which follows from Eq. (\ref{BPPowerEq})), this is an eigenstate of the plaquette projector with eigenvalue $1$; it also has no vertex violations, since $\hat{B}_P$ and $\oQ_V$ commute.   Hence 
\be
\langle \Psi_0|_{\Gamma=0}\hat{\mathcal{O}}_s   |\Psi_0 \rangle_{\Gamma=0} = 
\langle 0 | \left( \prod_P B_P \right )  \hat{\mathcal{O}}_s   \left(  \prod_P B_P \right )   |0\rangle
\ee
Since by definition the spin is $0$ on every edge in the trivial state $|0 \rangle$, it follows that this expectation value
 is completely determined by the coefficients that result from moving $\hat{\mathcal{O}}_s$ past the product over plaquette projectors.  These coefficients are necessarily the same in both models (see Eq. (\ref{EdgeJunk})).  Hence this expectation value must be identical in the \AWW and $\mathbb{Z}_{p}^\text{Potts}$ ground states.  
Intuitively, this is because 
 the matrix elements of $\hat{\Phi}_P$ and $\left( \hat{\Phi}_P \hat{\Theta}_{P,1}  \right)$ between any pair of spin configurations differ only by a phase; since different spin configurations are orthogonal, operators diagonal in the spin basis cannot measure these phases. 
 
 Notice that in the $\mathbb{Z}_{p}^\text{Potts}$ case, Eq. (\ref{GSMake}) describes only one of the possible ground state sectors in periodic boundary conditions.  Hence any statements that we can make about the correspondence in the phase transition between the two models are valid in the trivial ground state only.  This will not affect the thermodynamic properties of the transition, however.

Finally, we must show that the coefficients $a_\alpha^{(\Gamma/\lambda)}$ are the same for both models.  
This is clearly true to any order in perturbation theory: at $\Gamma =0$ the spectra of the two Hamiltonians (in the topologically trivial sector) are identical,  and the perturbing term $\Gamma \sum_e \oh_e$ has the same effect on both of these spectra.   Beyond the phase transition perturbation theory is no longer valid; however, we may re-express the wave-function (\ref{0Form}) as:
\be \label{0Form2}
|\Psi_0 \rangle = \sum_\alpha b_\alpha^{(\lambda/\Gamma )} \prod_{k \in \{ \tilde{\mathcal{C}}_\alpha \} }  \hat{B}_{P_k} |0 \rangle
\ee
 where $ \{ \tilde{\mathcal{C}}_\alpha \}$ is a set of plaquettes.  The coefficients $b_\alpha^{(\lambda/\Gamma )}$ must be equal in both models to any order in perturbation theory, for the same reasons as in the uncondensed phase.  
Further, if $e$ is an edge of $P$, in the $\mathbb{Z}_p^{\text{Potts}}$ case we have:
\ba \label{SwitchCoeffs1}
\oh_e \hat{B}_P& =&\frac{1}{p^2}  \sum_{n=0}^{p-1}  \left( \oS_{e, +} \right)^n \sum_{m=0}^{p-1} \left ( \hat{B}_{P,1} \right )^m \n
& =&\frac{1}{p^2} \sum_{m,n} e^{i  2 \pi m n/p} \left ( \hat{B}_{P,1} \right )^m  \left( \oS_{e, +} \right)^n 
\ea
(If $e$ is not an edge of $P$, the two operators commute). 
Hence 
\be \label{SwitchCoeffs2}
\oh_e \hat{B}_P |0 \rangle = \frac{1}{p^2} \sum_{m=0}^{p-1} \left ( \hat{B}_{P,1} \right )^m \sum_{n=0}^{p-1}  e^{i  2 \pi m n/p}  |0 \rangle  = \frac{1}{p} |0 \rangle 
\ee
Thus the wave function (\ref{0Form2}) can also be expressed as:
\be \label{LastEqn}
 \sum_\alpha b_\alpha^{(\lambda/\Gamma )} \prod_{k \in \{ \tilde{\mathcal{C}}_\alpha \} }  \hat{B}_{P_k} |0 \rangle = \sum_\alpha b_\alpha^{(\lambda/\Gamma )} \prod_{k \in \{\tilde{ \mathcal{C}}^*_\alpha \} }  \left( p \ \oh_{e_k} \right ) |\Psi_0 \rangle_{\Gamma =0}
 \ee
where $\{ \tilde{\mathcal{C}}^*_\alpha \}$ contains at least one edge of every plaquette {\it not} in  $ \{ \tilde{\mathcal{C}}_\alpha \}$ (and no edges of the plaquettes in  $ \{ \tilde{\mathcal{C}}_\alpha \}$).  (Including multiple edges of a particular plaquette will not affect the outcome).  
Since $b_\alpha^{(\lambda/\Gamma )}$ are equal in both models, this proves the desired result.  

If $\oh_e$ or $\hat{B}_P$ are not of the Potts form (in the sense that they do not contain all powers of $\oS$ or $\hat{\Phi}$ with equal amplitudes) Eqs, (\ref{SwitchCoeffs1}) and (\ref{SwitchCoeffs2}) no longer hold.  However, in Eq. (\ref{0Form}) we need not restrict $\oh_e$ to the transverse field operator in the Hamiltonian in order for the result to be valid; it suffices to express the ground-state wave function in terms of operators diagonal in the spin basis acting on the $\Gamma=0$ ground state.  We therefore replace Eq. (\ref{LastEqn}) with
\ba
 \sum_\alpha b_\alpha^{(\lambda/\Gamma )} \prod_{k \in \{ \tilde{\mathcal{C}}_\alpha \} }  \hat{B}_{P_k} |0 \rangle = \n
  \sum_\alpha b_\alpha^{(\lambda/\Gamma )} \prod_{j \in \{ \mathcal{C}_\alpha \} } \tilde{\bf{h}}_{e_j} \prod_{k \in \{\tilde{ \mathcal{C}}^*_\alpha \} }  \left( p \ \oh_{e_k} \right ) |\Psi_0 \rangle_{\Gamma =0}
 \ea
Here $\tilde{ \mathcal{C}}^*_\alpha$ is defined as above, and $\mathcal{C}_\alpha$ contains the remaining edges of the lattice (notably, it contains at least one edge of every plaquette in $\tilde{\mathcal{C}}_\alpha$).  Here $\tilde{\bf{h}}_{e} = \sum_{n=0}^{p-1} \alpha_n \left(\oS_{e,+}\right)^n$, with $\alpha_n$ chosen so that that
\be
\frac{1}{p}  \tilde{\bf{h}}_{e} \sum_{m=0}^{p-1} \hat{B}_{P,1}^m |0 \rangle =\hat{B}_P |0 \rangle
\ee
This gives an expression of the required form for the ground state at large $\Gamma/\lambda$, with coefficients that are necessarily the same in both models.

\subsection{SU(2)$_k$ Walker-Wang models}

Next we consider the set of transitions for SU(2)$_k$ in which spin $k/2$ vortex loops condense.  
The arguments we used in the Abelian case can essentially be applied here, as  they depend only on the commutation relations between $\hat{B}_{P,1}$ and the transverse field term, with a few modifications which we describe below.  

Specifically, if we shift the transverse field term by a constant, taking
\be
\oS_{e} = (-1)^{2 \os_e} \ \ \ \ \ \ \ \ \oh_e = \frac{1}{2} \left( 1 + \oS_e \right ) 
\ee
then $\oS_e$ and $\hat{\Phi}_{P, 1/2}$ anti-commute, satisfying Eq. (\ref{EdgeJunk}) with $p=2$.  Hence Eq.'s (\ref{0Form}) - (\ref{OExt}) apply, and we need only show that the expectation values of any product of transverse field terms in the unperturbed ground state, and the coefficients $a_{\alpha}^{(\Gamma/\lambda)}$ in (\ref{0Form}) are identical to those of the \gz gauge theory.  

For SU(2)$_k$,  Eq. (\ref{GSMake}) must be replaced by:
\be
|\Psi_0^{SU(2)_k} \rangle_{\Gamma =0} = \prod_P \frac{1}{2} \left(1 + \hat{\Phi}_{P,1}  - \sqrt{2} \hat{\Phi}_{P, 1/2} \right)  |0 \rangle
\ee
The transverse field term $(-1)^{2 \os_e}$ commutes with $1 + \hat{\Phi}_{P,1}$, but anti-commutes with $\hat{\Phi}_{P, 1/2}$.   Thus moving the transverse field past the plaquette operators has the identical effect as in the \gz gauge theory, where 
\be
|\Psi_0^{\mathbb{Z}_2}  \rangle_{\Gamma =0} = \prod_P \frac{1}{2} \left(1 + \hat{\Phi}_{P}  \right)  |0 \rangle
\ee
and $\oS_e$ commutes with $1$, and anti-commutes with $\hat{\Phi}_{P}$.  It follows, by the same reasoning as used in the Abelian case, that 
\ba \label{SameEh}
\langle \Psi_0^{SU(2)_k} |_{\Gamma=0}\hat{\mathcal{O}}_s^{(SU(2)_k)}   |\Psi_0^{SU(2)_k} \rangle_{\Gamma=0} \n
 = \langle \Psi_0^{\mathbb{Z}_2}|_{\Gamma=0}\hat{\mathcal{O}}_s^{(\mathbb{Z}_2)}   |\Psi_0 ^{\mathbb{Z}_2} \rangle_{\Gamma=0} 
\ea
where $\hat{\mathcal{O}}_s^{(SU(2)_k)}$ measures spin mod $1$ in the SU(2)$_k$ model, and $\hat{\mathcal{O}}_s^{(\mathbb{Z}_2)} $ measures the Ising spin of the \gz gauge theory.  
Similarly the arguments leading to Eq. (\ref{LastEqn}) remain valid, provided that we replace the trivial state $|0 \rangle$ with the state obeying (\ref{EqRestricted}).  This completes the proof.  

We emphasize that Eq. (\ref{SameEh}) does not hold if we include operators that can differentiate between integer spins in the SU(2)$_k$ models.   Such operators will not commute with $\hat{\Phi}_{P,1}$, and this component of the plaquette term (which is analogous to the identity component of the plaquette term in the \gz case) is not left invariant by moving the spin-measuring operators to the right.

As mentioned in the main text, this result is somewhat surprising in light of the fact that the matrix elements of $\hat{\Phi}_{P, 1/2}$ in the SU(2)$_k$ model and $\hat{\Phi}_{P}  $ in the \gz model differ both in sign and magnitude, and because a given configuration of half-integer spin loops in the SU(2)$_k$ model corresponds to many different spin configurations.   
It is instructive to understand why these differences do not affect the expectation values of any operator that satisfies  the conditions (\ref{EqRestricted}) in the $\Gamma =0$ ground states.  This is most apparent in the SU(2)$_2$ case, though the counting carries through for any $k$.\cite{Chandranetal}

As for the Abelian case described above, operators satisfying (\ref{EqRestricted}) are insensitive to the phases of the matrix elements.  This is clearly true for any operator diagonal in the spin basis, and also (by the arguments above) for $\hat{\Phi}_{P, 1/2}$. 
Less trivially, the difference in magnitude and multiplicity of configurations turn out to cancel each other: in a given spin configuration in the SU(2)$_2$ model, any contractible spin-$1/2$ loop can be removed and replaced with one of exactly two configurations of spins $0$ and $1$.  This is because (1) the number of spin $1$ edges external to any spin $1/2$ loop must be even if Eq. (\ref{EqRestricted}) is satisfied; and (2) for a given loop on which the total number of external spin $1$ variables is even, there are exactly two configurations of integer spins on this loop which satisfy the vertex condition everywhere.  The first condition follows from the fact (explained in more detail in Ref. \onlinecite{WWUs}) that for $\Gamma=0$ the ground state of a Walker-Wang model contains only configurations that are ``allowed" as link diagrams within the category; a diagram with a spin-1/2 loop and an odd number of spin-1 external edges is not such a diagram.   To see the second fact, we first note that if all edges external to the loop have spin 0, then we may replace the spin-1/2 loop with either a closed loop of spin 0 or of spin 1.  If there are at least two external edges of spin 1, then the two ways to assign integer spins to the loop correspond to beginning at one of the spin-1 external legs, and assigning a spin of 1 to one of the two edges on the loop with which it shares a vertex.  After this initial choice there is no more freedom in the spin assignments, provided that the spins of all external edges are fixed and we do not introduce any vertex violations.  

By the above reasoning, if we add  (or remove) a closed loop if spin-1/2 edges, we halve ( or double) the number of possible integer spin configurations compatible with the given choice of spin-1/2 loops.  However, this is exactly compensated for by the fact that the coefficients in the action of the plaquette projector multiply such configurations by a factor of $\sqrt{2}$ (or $1/\sqrt{2})$ relative to the state from which they were derived.  Hence the {\it probability} to be in a given loop configuration is the same in both models.  
Hence from the point of view of operators only sensitive to the spin mod $1$, the ground-state loop configurations of the SU(2)$_2$ Walker-Wang model look exactly like the loop configurations in the ground state of the \gz model.  

Needless to say, here again the arguments  apply only to the trivial ground-state sector of the \gz theory; the other ground-state sectors (where they exist) do not have analogues in the Walker-Wang case.  

\section{More general Abelian Walker-Wang models and their transitions} \label{complicatedWWmodels}
In \secref{AWWSec} we considered \AWW models, which comprise only a subset of all abelian Walker-Wang models. In this section we examine vortex transitions in more general abelian Walker-Wang models.  Since the arguments of Appendix \ref{PhaseTranApp} can be used to map the phase transitions onto those of discrete Abelian gauge theories, we will focus here on the nature of the condensed phase.  It will be useful to speak about Walker-Wang models in terms of their corresponding categories. The $p$-state \AWW should be thought of as being based on the category $\integ^{1/2}_p$ for $p$ even, and $\integ^{(p-1)/2}_p$ for $p$ odd\cite{BondersonThesis}. Other examples include the abelian Potts gauge theories of \secref{PottsSec}, which can be viewed as Walker-Wang models based on $\integ^{0}_p$. 

In general, $p$-state abelian Walker-Wang models are based on categories $\integ^{q}_p$, where $0\leq q\leq p-1$ is \emph{either integer or half-integer}, and where $q$ can only ever be half integer if $p$ is even. These categories are described  in Ref. \onlinecite{BondersonThesis}.  The labels in $\integ^{q}_p$ are $\{0,1,\ldots,p-1\}$. The general $M$-matrix of category $\integ^{q}_p$ takes the form $M_{ab}=e^{i 4\pi q a b/p}$, and as before we can use this to understand the topological order of the corresponding Walker-Wang model -- remember that if $M$ has $c$ columns filled with $1$, then the corresponding Walker-Wang model has $c$ deconfined species, and a topological degeneracy of $c^{3}$ on the 3-torus. $c = \gcd(2q,p)$ in the case of $\integ^{q}_p$.

Condensing vortices with label $m$ out of the $\integ^{q}_p$ Walker-Wang model leads to a phase described by a $\integ^{q p/g}_g$ Walker-Wang model, where $g=\gcd\db{2 q m,p}$. To see why, notice that the labels remaining in the condensed phase are precisely those $a$ satisfying $e^{i 4\pi q a m/p}=1$. The most general values of $a$ satisfying this equation comprise the set $\{0,1,2,\ldots,g-1\}\times p/g$, and so precisely $g$ labels remain in the condensed phase. 

The new model inherits the old fusion and braiding rules, as well as the $M$-matrix from $\integ^{q}_p$, but restricted to the $g$ new labels. Indeed, the new model can be thought of as being based on the category $\integ^{qp/g}_g$, which has labels $x\in\{0,1,\ldots,g-1\}$ related to the old labels by $a = x p/g$. We can understand the topological properties of this phase by examining the new $M$-matrix which takes the form $\widetilde{M}_{xy}= e^{i 4\pi \frac{qp}{g} x y / g}$. While the un-condensed model had $c= \gcd(2q,p)$ deconfined labels, the condensed model has $\widetilde{c}=\gcd(2qp/g,g)$ of them. In the $\integ_p$ gauge theories (which have $q=0$), $c=p$ while $\widetilde{c}=g$.  In the \AWW models, there is one deconfined label ($a=0$), while $\tilde{c} = \text{gcd} ( p/g, g )$ for $p$ even, and $ \text{gcd} ( p(p-1)/g, g ) =  \text{gcd} ( p/g, g )$ for $p$ odd.

\section{Fermions versus bosons in Walker-Wang models} \label{FermiApp}
	\begin{figure}
\begin{center}
\includegraphics[width=.975\linewidth]{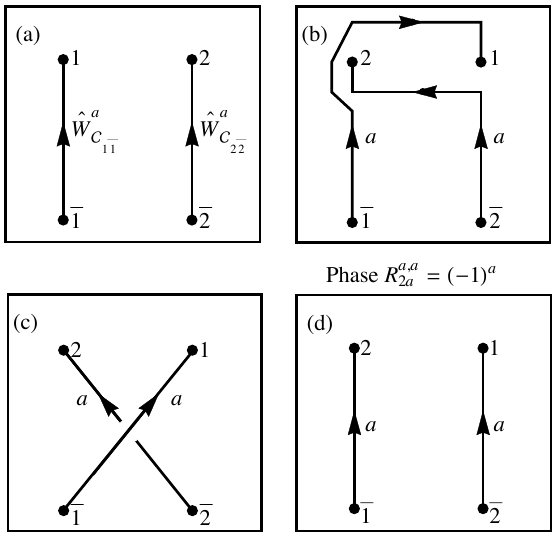}\\
                \caption{(a) shows point sources in an abelian Walker-Wang model with ``Fermionic excitations'', where $1,2$ have charge $a$ and $\ov{1},\ov{2}$ have charge $-a$. The strings represent the operators that act on the ground state to form these point excitations. (b) Shows the diagram resulting from the exchange of the defects. Plaquettes operators can be used to hop the string to the configuration in (c). The over-crossing in (c) can be removed at the cost of introducing a possible minus sign, and thus the diagram can be deformed back to (a) but with a phase of $(-1)^a$, which gives a fermionic or bosonic statistic contingent on whether $a$ is odd or even respectively.}
                \label{exchange}
\end{center}                
\end{figure}

In \secref{GenAWWSect} we claimed that spin labels $a$ for which the eigenvalue of $\hat{\Theta}_{P,a}$ can be only $\pm 1$ are associated with fermionic ``charge" excitations, while if $\hat{\Theta} \equiv 1$ then the associated charge is a boson. In the language of categories this is the statement that Walker-Wang models based on $\integ^{p/2}_p$ (which can only be defined for even $p$) have deconfined fermionic and bosonic excitations, while those based on $\integ^{0}_p$ have all their point-like excitations deconfined and bosonic. In this section we show how fermionic statistics arise in Walker-Wang models based on $\integ^{p/2}_p$.

To begin, act on the ground state with a deconfined string operator $\hat{W}^a_{\maC_{1,\ov{1}}}$ which connects vertices $1$ to $\ov{1}$, and creates two defects with conjugate charges $a,\ov{a}\in\{0,1,\ldots,p-1\}$ (note $a\neq \ov{a}$ in general). Similarly use $\hat{W}^a_{\maC_{2,\ov{2}}}\ket{GS}$ to create an identical pair of defects of charge $a,\ov{a}$ at $2$ and $\ov{2}$ respectively, thus producing the excited state $\hat{W}^a_{\maC_{1,\ov{1}}}\hat{W}^a_{\maC_{2,\ov{2}}}\ket{GS}$ (see \figref{exchange}(a)). The prescription for forming string operators is discussed in detail in Ref~\onlinecite{WWUs}. In the cases of $\integ^{p/2}_p$ and $\integ^{0}_p$, the string operator creates deconfined excitations because it commutes with the plaquette operators along its length; this is possible because $\integ^{p/2}_p$ has a trivial $M$-matrix\cite{WWUs} $M_{cd}\equiv 1$. 

The exchange operation is only meaningful for identical particles, thus we begin by exchanging sources $1$ and $2$ (which have equal charges) which corresponds with adiabatically evolving the state \figref{exchange}(a) to the one in \figref{exchange}(b). This new state (a) will be equivalent to (b), except for an overall phase. To evaluate this phase, we first note that if $P$ is a plaquette touching a path $\maC$, then
\be 
(\hat{\Phi}_P)^a\hat{\Theta}_{P,a}\hat{W}^a_{\maC}  \ket{GS} = \hat{W}^a_{\maC'}\ket{GS}
\ee
where $\maC'$ is the same path as $\maC$ except it takes a detour around plaquette $P$ (this step relies on the fact that $\integ^{p/2}_p$ has trivial $M$ matrix). We can now use the fact that $\hat{\Theta}_{P,a}=1$, and the fact this operator commutes with $\hat{W}^a_{\maC}$, to shown that $\hat{W}^a_{\maC'}$ and $\hat{W}^a_{\maC}$ have the same effect on the ground state.  Using this idea repeatedly, it is easy to show that \figref{exchange}(b) and \figref{exchange}(c) represent the same state. 

Notice that the state in \figref{exchange}(c) has an over-crossing of the string operators (from the point of view of the projection). The overcrossing can be removed and we can connect $1$ to $\ov{2}$ and $2$ to $\ov{1}$ to form \figref{exchange}(d), but there will be a phase associated with this process, as illustrated by
\be
\mathord{\includegraphics[height=14ex]{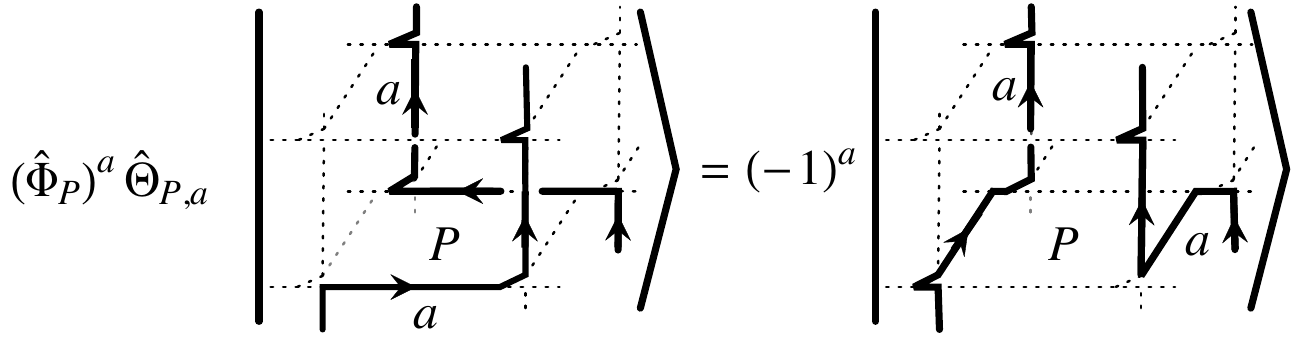}}\punc{,}
\ee
which shows that $(\hat{\Phi}_P)^a\hat{\Theta}_{P,a}$ introduces a phase if it changes the number of over-crossings in a configuration. Notice that $(\hat{\Phi}_P)^a \hat{\Theta}_{P,a}=1$ on the ground state, so configurations that are related by removing such an over crossing have a relative sign of $(-1)^a$ in the ground state superposition. One can extend this reasoning to show that the state in \figref{exchange}(c) is the same as \figref{exchange}(d), except for this relative phase. Thus, the exchange of the two defects has resulted in a configuration identical to that in \figref{exchange}(a), but with a phase of $(-1)^a$. 

One can more carefully derive this exchange phase by noting that, when a category has trivial $M$-matrix like $\integ^{p/2}_p$, the strings representing operators can be deformed according to the graphical rules of the category seen in \figref{fig:gencat}. This fact follows from the `handle-slide' property\cite{WWUs,WW} of Walker-Wang models. To get between \figref{exchange}(c) and (d), one may use the manipulations of the string operators
\be
\mathord{\includegraphics[height=13ex]{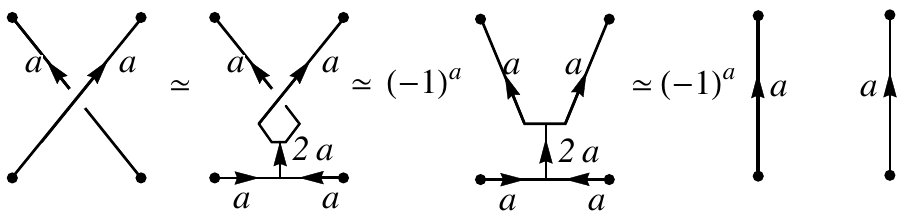}}\punc{.}
\ee
which correspond with a use of the rules in \figref{fig:gencat}, with \figref{fig:gencat}(e) needed specifically for the second equality, where we used the fact that $R^{a\ov{a}}_{2a}=(-1)^a$ for $\integ^{p/2}_p$. The statistic arises precisely because the strings connecting the point defect keep track of the exchange.

\bibliography{WWCondBib}

\end{document}